\documentclass[10pt,journal,compsoc]{IEEEtran}

\usepackage{amsmath}
\usepackage{wrapfig}
\usepackage{multirow}
\usepackage{amsmath}
\usepackage{wrapfig}
\usepackage{enumerate}
\usepackage{color}
\usepackage{colortbl}

\usepackage{hhline}
\usepackage[ruled,linesnumbered]{algorithm2e}
\usepackage{mathtools}
\usepackage{makecell}

\usepackage{times}
\usepackage{epsfig}
\usepackage{graphicx}
\usepackage{amssymb}
\usepackage{bm}
\usepackage{comment}
\usepackage{color,xcolor}
\usepackage{xspace}
\usepackage{subfigure}
\usepackage{setspace}

\usepackage{booktabs} % For formal tables
\usepackage[ruled]{algorithm2e} % For algorithms

\definecolor{airforceblue}{rgb}{0.36, 0.54, 0.66}
\definecolor{arsenic}{rgb}{0.23, 0.27, 0.29}

\newcommand{\bfB}{\mathbf{B}}
\newcommand{\bfb}{\mathbf{b}}
\newcommand{\bfc}{\mathbf{c}}
\newcommand{\bfC}{\mathbf{C}}
\newcommand{\bfD}{\mathbf{D}}
\newcommand{\bfd}{\mathbf{d}}
\newcommand{\calD}{\mathcal{D}}
\newcommand{\calE}{\mathcal{E}}
\newcommand{\bfF}{\mathbf{F}}
\newcommand{\bfI}{\mathbf{I}}

\newcommand{\bfn}{\mathbf{n}}

\newcommand{\bfR}{\mathbf{R}}
\newcommand{\bft}{\mathbf{t}}
\newcommand{\bfu}{\mathbf{u}}
\newcommand{\bfv}{\mathbf{v}}
\newcommand{\bfw}{\mathbf{w}}
\newcommand{\bfx}{\mathbf{x}}
\newcommand{\bbX}{\mathbb{X}}
\newcommand{\bfy}{\mathbf{y}}
\newcommand{\bbY}{\mathbb{Y}}
\newcommand{\bmalpha}{\bm{\alpha}}
\newcommand{\bmbeta}{\bm{\beta}}

\ifCLASSOPTIONcompsoc
  \usepackage[nocompress]{cite}
\else
  \usepackage{cite}
\fi
\ifCLASSINFOpdf
\else
\fi
\begin{document}
\title{3D Magic Mirror: Automatic Video to 3D Caricature Translation}

\author{
	Yudong Guo$^\star$,~~~
	Luo Jiang$^\star$,~~~
	Lin Cai,~~~
	Juyong Zhang$^\dagger$
\IEEEcompsocitemizethanks{\IEEEcompsocthanksitem Y. Guo, L. Jiang, L. Cai and J. Zhang are with School of Mathematical Sciences,
	University of Science and Technology of China.}
\thanks{$^\star$ indicates equal contribution.}\\
\thanks{$^\dagger$Corresponding author. Email: \texttt{juyong@ustc.edu.cn}.}
}

\IEEEtitleabstractindextext{%
\begin{abstract}
Caricature is an abstraction of a real person which distorts or exaggerates certain features, but still retains a likeness. While most existing works focus on 3D caricature reconstruction from 2D caricatures or translating 2D photos to 2D caricatures, this paper presents a real-time and automatic algorithm for creating expressive 3D caricatures with caricature style texture map from 2D photos or videos. To solve this challenging ill-posed reconstruction problem and cross-domain translation problem, we first reconstruct the 3D face shape for each frame, and then translate 3D face shape from normal style to caricature style by a novel identity and expression preserving VAE-CycleGAN. Based on a labeling formulation, the caricature texture map is constructed from a set of multi-view caricature images generated by CariGANs~\cite{CaoLY18}. The effectiveness and efficiency of our method are demonstrated by comparison with baseline implementations. The perceptual study shows that the 3D caricatures generated by our method meet people's expectations of 3D caricature style.
\end{abstract}

}

\maketitle

\IEEEdisplaynontitleabstractindextext

% \fi
\IEEEpeerreviewmaketitle

\IEEEraisesectionheading{\section{Introduction}}
\label{sec:intro}

% What's the problem? real-time 3D caricature generation and performance capture
% Why is it important and its applications.

\IEEEPARstart{A} caricature is a description of a person in a simplified or exaggerated way to create an easily identifiable visual likeness with a comic effect~\cite{Sadimon10survey}. This vivid art form contains the concepts of exaggeration, simplification and abstraction, and has wide applications in cartoon characters, custom-made avatars for games and social media. Although experienced artists could draw a 2D caricature without losing its distinct facial features with a given photo, it is still not easy for common people. Motivated by this, there have been a lot of work by designing computer-assisted systems to generate caricatures with few user inputs~\cite{Brennan85,liang2002example,Shum-CVPR03,liao2004automatic,CGF:CGF1804,han2017deepsketch2face,Han2019Caricature,WuZLZC18} or automatically with a reference caricature~\cite{CaoLY18}. Most of these works focus on 2D caricatures generation or 3D caricatures modeling from user inputs such as facial landmarks and sketches.

%In this paper, we develop techniques for real-time, living and personal specific 3D caricature performance capture with a monocular camera.

%% what is the problem and the main challengings of 2D photos to 3D caricatures
Drawing some simple exaggerated faces like big noses and small eyes is easy for common people, while very few people can draw sketches of caricature which capture personal-specific characteristics of the subject from others. Furthermore, we are more interested in exaggerating a given photo or video with controllable styles in the 3D animation world. To generate a 3D animation sequence by user interaction is a repetitive and tedious work. Therefore, automatic 3D personalized caricature generation from 2D photos or videos is more meaningful and useful. However, none of existing works considers this problem before. On the other hand, 3D caricature modeling from 2D photo is not only an ill-posed reconstruction problem but also a cross-domain translation problem. A straightforward strategy is to first translate the 2D photos to 2D caricatures, and then reconstruct 3D caricature models from 2D caricatures. However, automatic 3D reconstruction from 2D caricature is not easy due to the diversity of caricatures.

\begin{comment}
which is widely studied in computer graphics and computer vision. In order to eliminate the variations on geometric and texture, model based methods first build a low-dimensional parametric representation of 3D face models from a 3D face data set and then fit the parametric model to the input 2D photos~\cite{blanz1999morphable,cao2014facewarehouse,tewari18FaceModel}. The geometry details can be recovered by shape-from-shading approach~\cite{kemelmacher20113d,kemelmacher20083d,jiang20173d}. Another related problem is to directly translate 2D photos to 2D caricatures, which is an image-to-image translation problem and is a hot topic in recent years~\cite{ZhuPIE17,ZhuZPDEWS17}. Due to the large gap of shape and appearance between photos and caricatures, a single network can not handle this translation well. Very recently, \cite{CaoLY18} proposes to decompose the challenging photo-to-caricature translation into two tasks, one for geometric exaggeration and another for appearance stylization.
\end{comment}

%%The following part needs revision.
Although much progress has been made in 2D and 3D caricatures generation, there still exist several challenges for automatic 3D caricatures generation from 2D photos. First, different with interactive based methods which can use landmarks or sketches to express exaggerated shapes, the input 2D photos or videos only contain information for regular 3D face reconstruction, not in an exaggerated way. Moreover, the diversity of caricatures is much more severe than regular faces, which means that model based methods for 3D regular face reconstruction can not be simply extended to 3D caricature modeling. Second, even an experienced artist would take a long time to create a 3D caricature that allows people to easily recognize its identities. This is because it is not easy to define a similarity measurement between regular face models and exaggerated face models. It becomes even harder for a 3D caricature sequence with different expressions to maintain the same identity. Third, although we can easily get a caricature style image with method like~\cite{CaoLY18}, it is not a trivial task to generate a complete caricature style texture map due to the inconsistency between caricatures from different views.

%% how we solve this problem. This paragraph needs revision. How we solve these problmes?
In recent years, 3D facial performance capture from a monocular camera has achieved great success~\cite{bouaziz2013online,cao2014facewarehouse,ichim2015dynamic,garrido2016reconstruction} based on parametric models like 3DMM~\cite{blanz1999morphable}, FaceWareHouse~\cite{cao2014facewarehouse} and FLAME~\cite{FLAME:SiggraphAsia2017}. Therefore, we first reconstruct the 3D face shape for each frame of the input video, and then apply geometry-to-geometry translation on 3D face shape from regular style to caricature style with graph-based convolutional neural network. To automatically translate the 3D face shape from regular style to caricature style, we first train variational autoencoders on our well-designed data set to encode 3D regular face and 3D caricature face separately in the latent space. An identity and expression preserving cycle-consistent GAN for mapping between 3D regular face and 3D caricature face is applied on the trained latent space to translate between these two styles. To generate a caricature texture map, we first apply a caricature style appearance generation method to a user's multi-view photos, and then fuse different views' appearance together based on a novel labeling framework. All components of our algorithm pipeline run automatically, and it runs at least 20Hz. Therefore, our method could be used for 3D caricature performance capture with a monocular camera.

In summary, the contributions of this paper include the following aspects.
\begin{itemize}	
	\item We present an automatic and real-time algorithm for 3D caricature generation from 2D photos or videos, where geometry shape is exaggerated and caricature style is transferred from caricatures to photos. To the best of our knowledge, this is the first work about automatic 3D caricature modeling from 2D photos.
	\item We present a caricature texture map generation method from a set of multi-view photos based on a novel labeling framework. An identity and expression preserving translation network is presented to convert the 3D face shape from regular style to caricature style.
	\item We construct a large database including 2D caricatures and 3D caricature models to train the 2D caricature generation model and geometry-to-geometry translation model. The landmarks for each 2D caricature are also included. The database will be made publicly available.
\end{itemize}

\section{Related Work}
\label{sec:related-work}

According to the algorithm input and output, we classify the related works into the following categories.

% write about 3DMM, facewarehouse and some face database, briefly
\noindent{\textbf{3D Face reconstruction.}} 3D face reconstruction from photo, monocular RGB and RGB-D cameras are well studied~\cite{bouaziz2013online,CaoHZ14,thies2015realtime,guo20173d,jackson2017large,jiang20173d,tewari18FaceModel} in recent years, and \cite{Zollhoefer2018FaceSTAR} gives a complete survey on this topic. Due to high similarities between 3D faces, 3D face reconstruction methods always adopt data-driven approaches. For example, Blanz and Vetter proposed a 3D morphable model (3DMM)~\cite{blanz1999morphable} that was built on an example set of 200 3D face models describing shapes and textures. Later, \cite{vlasic2005face} proposed to perform multi-linear tensor decomposition on attributes including identity, expression and viseme. Based on parametric models, Convolutional Neural Networks (CNNs) were constructed to regress the model parameters and thus 3D face models are reconstructed~\cite{jackson2017large,tewari17MoFA}. \cite{cao2014facewarehouse} used RGB-D sensors to develop FaceWareHouse, which is a bilinear model containing 150 identities and 47 expressions for each identity. Based on FaceWareHouse, \cite{CaoHZ14,jiang20173d} regressed the parameters of the bilinear model of \cite{vlasic2005face} to construct 3D faces from a single image.

\noindent{\textbf{2D Face Caricatures.}} Since the seminal work~\cite{Brennan85}, many attempts have been made to develop computer-assisted tools or systems for creating 2D caricatures. \cite{akleman1997making,GoochRG04} developed an interactive tool to make caricatures using deformation technologies. \cite{liang2002example} directly learned the rules on how to change the photos to caricatures by a paired photo-caricature dataset. \cite{ShetLEC05} trained a cascade correlation neural network to learn the drawing style of an artist, and then applied it for automatic caricature generation. \cite{liao2004automatic} developed an automatic caricature generation system by analyzing facial features and using one existing caricature image as the reference. \cite{CaoLY18} proposed a GAN based method for photo-to-caricature translation, which models geometric exaggeration and appearance stylization using two neural networks. Different from the GAN based method in~\cite{CaoLY18}, ~\cite{Han2019Caricature} first performed exaggeration according to the input sketches on the recovered 3D face model, and then the warped image and re-rendered image are integrated to produce the output 2D caricature.

\noindent{\textbf{3D Face Caricatures.}} Relatively there is much less work on 3D caricature generation~\cite{o1997three,o19993d}. \cite{clarke2011automatic} proposed an interactive caricaturization system to capture deformation style of 2D hand-drawn caricatures. The method first constructed a 3D head model from an input facial photograph and then performed deformation for generating 3D caricatures. Similarly, \cite{VieiraVN13} applied deformation to the parts of the input model which differ most from the corresponding parts in the reference model. \cite{liu2009semi} proposed a semi-supervised manifold regularization method to learn a regressive model for mapping between 2D real faces and the enlarged training set with 3D caricatures. \cite{SelaAK15} exaggerated the input 3D face model by locally amplifying its area based on its Gaussian curvature. \cite{WuZLZC18} formulated the 3D caricature modeling as a deformation problem from standard 3D face datasets. By introducing an intrinsic deformation representation which has the capability of extrapolation, exaggerated face model can be produced while maintaining face constraints with the landmark constraint. With the development of deep learning, \cite{han2017deepsketch2face} developed a sketch system using a CNN to model 3D caricatures from simple sketches. In their approach, the FaceWareHouse~\cite{cao2014facewarehouse} was extended with 3D caricatures to handle the variation of 3D caricature models since the lack of 3D caricature samples made it challenging to train a good model. Different from these existing works, our approach directly translates the input 2D photos to 3D caricatures, which has never been considered before and is more challenging.

\noindent{\textbf{Style Transfer using GAN.}} Recently, Generative Adversarial Nets (GAN)~\cite{GoodfellowPMXWOCB14} has been widely used in style transfer by jointly training a generator and a discriminator such that the synthesized signals have characteristics indistinguishable from the target sets. With paired training set, \cite{IsolaZZE17} proposed pix2pix network with a conditional GAN for applications like photo-to-label, photo-to-sketch. For unpaired training set, \cite{ZhuPIE17} proposed CycleGAN via a novel cycle-consistency loss to regularize the mapping between a source domain and a target domain. Recently, CycleGAN has been successfully applied to deformation transfer between 3D shape sets~\cite{GaoYQLRXX18}. Different from the original CycleGAN, their VAE-CycleGAN applies cycle-consistency loss on the latent spaces. \cite{KimCTXTNPRZT18} presented a GAN based approach for face reenactment by transferring the full 3D head information from a source actor to a portrait video of a target actor. To generate photo-realistic facial animations with a single-view portrait photo, \cite{GengSZWZ18} proposed to first wrap the photo by 2D facial landmarks and then synthesize the fine-scale local details and hidden regions by GAN models.
\section{Overview}
\label{sec:overview}
\begin{figure*}[t]
\begin{center}
\includegraphics[width=1\linewidth]{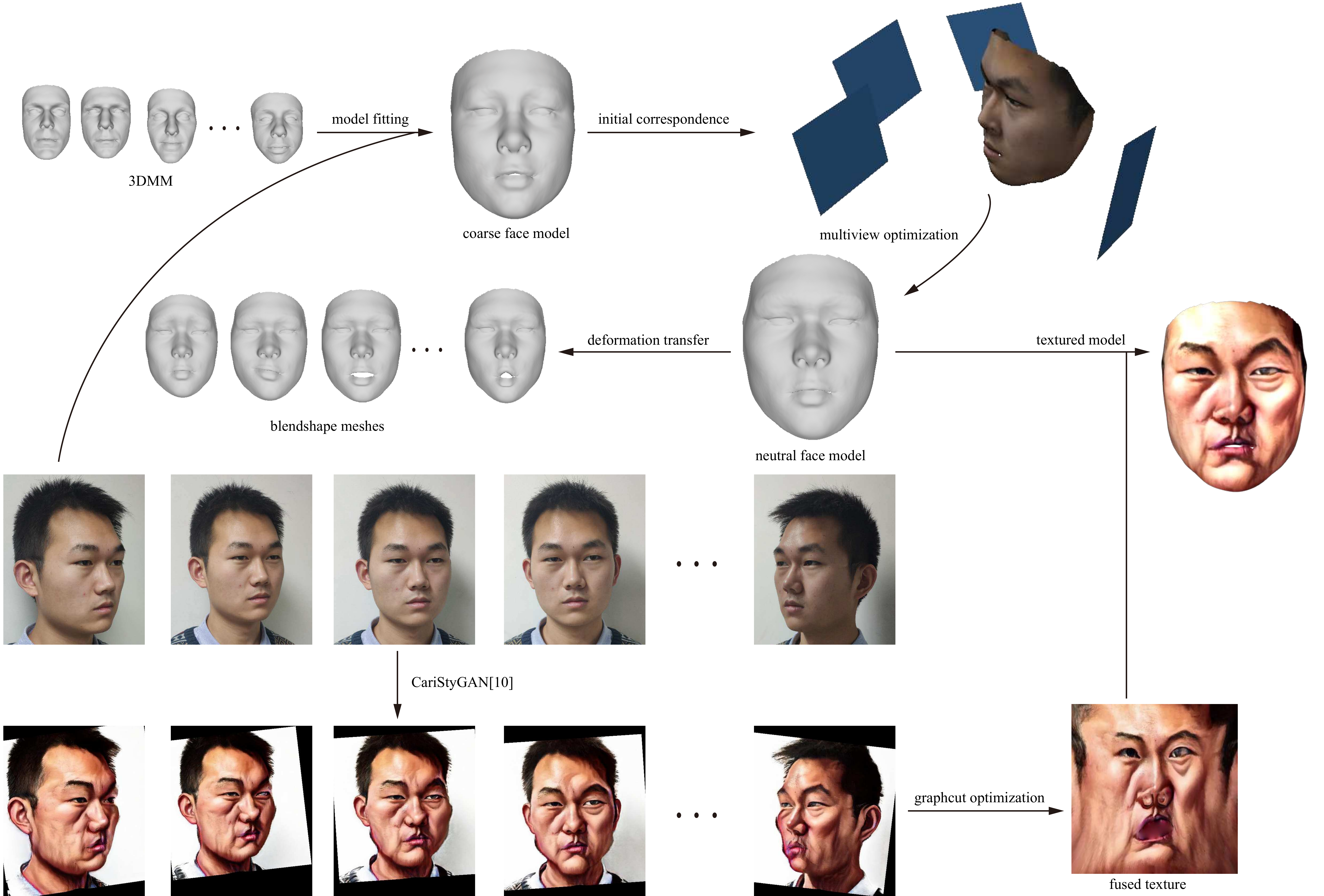}
\end{center}
   \caption{Given a set of multi-view facial images, we first fit a coarse parametric face model to all input images, and refine the coarse face model with a multiview method to generate the neutral face model $\bfb_{0}$. Other 46 blendshape meshes are constructed by deformation transfer from the neutral face model. Caricature texture maps are generated by the input images and are fused by a graph cut based optimization method.}
\label{fig:static_model}
\end{figure*}

We propose a two stages based method for 3D caricature performance capture from a monocular camera in real-time. In the first static modeling stage, we collect a set of photos with neutral expression from different views for the user. These photos are first used to construct a high-quality 3D face model by a novel multi-view optimization, and then a set of blendshapes are constructed by deformation transfer~\cite{sumner2004deformation}. In addition to reconstructing the geometry shape, we also generate a caricature texture map with a given reference caricature or random styles by fusing the multi-view caricatures based on a novel labeling framework. The algorithm pipeline of this part is given in Fig.~\ref{fig:static_model}, and the algorithm details are given in Section~\ref{sec:stactic}.

In the second dynamic modeling stage, we first build a 3D face tracking system to reconstruct the 3D regular face shape for each frame based on the blendshape constructed in the first stage. We then train a 3D face VAE-CycleGAN model to translate the 3D face shape from regular style to caricature style, such that the identities of generated 3D caricatures remain unchanged during tracking process and the expressions of the generated 3D caricature and the original face shape are similar for each frame. To make the generated 3D caricature animation sequence smooth, the generated 3D caricature is smoothed in the latent space. With the caricature texture map generated in static modeling stage, our system outputs a real-time, living and personal specific 3D caricature animation sequence driven by an actor and a monocular camera. The algorithm details on this part are given in Section~\ref{sec:dynamic}.

\section{Static Modeling}
\label{sec:stactic}
At the beginning, we first construct the blendshape for the user and the caricature style texture map with a set of photos in neutral expression from different views for the user. The algorithm pipeline of this part is given in Fig.~\ref{fig:static_model}, and the algorithm details of each part are detailed below.

\subsection{Blendshape Construction}
\label{sssec:blendshape}
In our regular face tracking system, we adopt the blendshape representation~\cite{LewisARZPD14} for real-time performance capture. The dynamic expression model is represented as a set of blendshape meshes $\bfB = [\bfb_0, \ldots ,\bfb_n]$, where $\bfb_0$ is the neutral expression and $\bfb_i$, $i > 0$ are a set of specific facial expressions. In order to construct the blendshape mesh $\bfb_0$ for the user, we acquire a set of uncalibrated images of the user in neutral expression from different views. Then a two-stage method is used to reconstruct a discriminative 3D face model. First, we build a coarse face model by fitting a parametric face model such as 3D Morphable Model (3DMM) to all the input $N$ facial images to minimize a joint objective energy that contains landmark alignment, photo consistency and regularization, see details in~\cite{garrido2016reconstruction}. In the first stage, we simulate the global illumination using the second order spherical harmonics (SH) basis functions
\begin{equation}
\bfI(\bfn_i,\rho_i|\gamma) = \rho_i\sum_{j = 1}^{9}\gamma_j\phi_j(\bfn_i),
\label{eq:rendering}
\end{equation}
where $[\phi_1(\bfn_i), \ldots, \phi_{9}(\bfn_i)]$ is the SH basis computed with the vertex normal $\bf{n}_i$, and $[\gamma_1,\ldots, \gamma_{9}]$ are the SH coefficients, $\rho_i$ is the gray-scale albedo value of the $i$-th vertex. The purpose of model fitting is to estimate the initial face geometry, pose of the input images, lighting and a global albedo map, see Fig.~\ref{fig:static_model}. The initial face shape is quite smooth and can not reconstruct the user's personal characteristics well limited by the representation ability of the parametric model.

In the following, the 3D face shape is further optimized to be more discriminative to identities. Based on the initial correspondences between images supplied by the reconstructed coarse face shape, we project the visible part of the coarse face model onto each view image and optimize a 2D displacement field $\Delta \bfu_i$ for these projections such that different view projections of the visible vertex have the same pixel value
\begin{equation}
E_{\textrm{dis}}(\Delta \bfu_i) = \sum_{j = 1}^N \lambda_{i}^{j}(\|\bfI_j(\bfu_i^j + \Delta\bfu_i^j) - \bfI(\bfu_i)\|_2^2 + \lambda_{\textrm{reg}}\|\Delta \bfu_i^j\|_2^2),
\label{eq:correspondence}
\end{equation}
where $\Delta \bfu_i=[\Delta\bfu_i^1, \ldots, \Delta\bfu_i^N]$ is a 2D displacement vector for the $i$-th vertex. $\bfI_{j}(\bfu)$ is the pixel color at the position $\bfu$ at in the $j$-th view image. $\bfI(\bfu_i)$ is the pixel color corresponding to the $i$-th vertex in the best view. The best view means that the angel between normal direction and view direction for the $i$-th vertex is smallest. $\lambda_{i}^{j}$ is set to 1 if the $i$-th vertex is visible in the $j$-th view, 0 otherwise. $\lambda_{\textrm{reg}}$ is a user-specified parameter. At last we optimize the sum of correspondence energy $E_{\textrm{dis}}(\Delta \bfu_i)$ at each visible vertex by Gauss-Newton method. With the optimized dense correspondences between all different view photos, we solve a bundle adjustment problem to jointly optimize a point cloud $\{\bfv_i\}$, the camera internal parameters $\Pi$ and poses $\{(\bfR_j, \bft_j)\}$ for each photo with the following energy:
\begin{equation}
\label{eq:bundle}
E = \sum_{i =1}^M\sum_{j=1}^N\|\Pi(\bfR_j \bfv_i + \bft_j) - (\bfu_i^j + \Delta\bfu_i^j)\|^2,
\end{equation}
where $M$ is the number of mesh vertices. And then we deform the coarse 3D face mesh to fit the optimized point cloud by making each vertex close to its nearest point using a Laplacian deformation algorithm~\cite{botsch2008linear}.

After obtaining the neutral face mesh $\bfb_0$, we construct the other 46 blendshape meshes ($\bfb_1,\ldots,\bfb_{46}$) for the user by applying deformation transfer~\cite{sumner2004deformation} from the FaceWareHouse~\cite{cao2014facewarehouse} expressions to $\bfb_0$. In this way, we construct a set of blendshape meshes $\bfB$ for each user.

%\begin{figure}[t]
%\begin{center}
%\includegraphics[width=1.0\linewidth]{pdfs/normal_texture.pdf}
%\end{center}
%   \caption{}
%\label{fig:normal_texture}
%\end{figure}

\subsection{2D Caricature Generation}
\label{2d_cari_gen}
Our aim is to reconstruct a 3D exaggerated face model with caricature style texture map. We directly adopt the state-of-the-art method~\cite{CaoLY18} to generate 2D caricatures for the multi-view image set acquired in the Sec.~\ref{sssec:blendshape}. To construct the caricature style texture map, we only need to change the style appearance from photo to caricature without distorting the geometry. To achieve the unpaired photo-to-caricature translation, we first construct a large-scale training dataset. For the photo domain, we randomly sample about 8k face images from the CelebA database~\cite{liu2015deep} and detect the 68 landmarks for each facial image using the method in~\cite{dlib09}. For the caricature domain, we select about 6k hand-drawn portrait caricatures from Pinterest.com and WebCaricature dataset~\cite{huo2017webcaricature} with different drawing styles and manually label 68 landmarks for each caricature. The face region of collected photos and caricatures are cropped to $512 \times 512$ by similarity transformation. After constructing the training data set, we directly follow the method proposed in~\cite{CaoLY18} to train the \textit{CariGeoGAN} and \textit{CariStyGAN} for 2D geometry translation and style translation separately. We train the \textit{CariGeoGAN} to learn geometry-to-geometry translation from photo to caricature, and warp~\cite{cole2017synthesizing} each caricature with its original landmarks to a new caricature with the landmarks translated by the \textit{CariGeoGAN}. Finally, we train the \textit{CariStyGAN} to learn appearance-to-appearance translation from photo to new caricature while preserving its geometry. For more details on this part, please refer to~\cite{CaoLY18}. Based on the trained \textit{CariStyGAN} model, we can translate our 2D face photos to caricatures with a reference style or a random style as shown in the last two rows of Fig.~\ref{fig:static_model}.

\subsection{Caricature Texture Map}
\label{texture_fusion}

\begin{figure}[t]
\begin{center}
\includegraphics[width=1.0\linewidth]{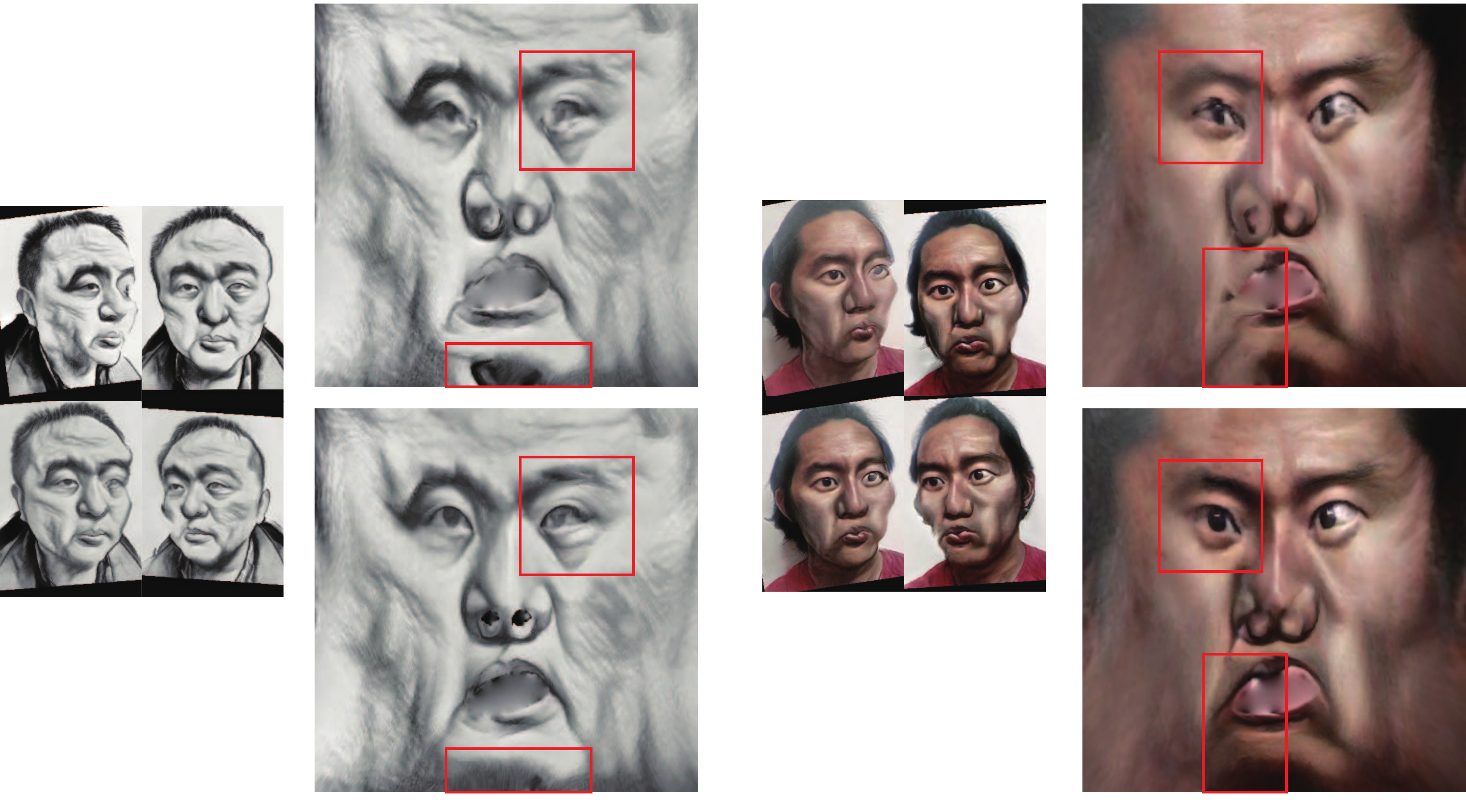}
\end{center}
   \caption{The complete caricature texture map from multi-view images. The figures in first row are the results only considering Eq.~\eqref{eq:viewsimilarity}, and the figures in second row are the results by our method.}
\label{fig:curicture_texture}
\end{figure}

Given the reconstructed neutral face mesh $\bfb_0$ and caricatures from different views generated by \textit{CariStyGAN}, the next step is to generate a caricature style texture map. We apply the latest UV parameterization~\cite{ShtengelPSKL17} to the face mesh to obtain the UV textures of different views. Different from facial images captured from different views, the generated 2D caricatures by \textit{CariStyGAN} might be inconsistent in different views. Therefore, fusing them together by selecting the color gradients of the pixels with the most parallel view rays to the surface normals as adopted in~\cite{ichim2015dynamic} for facial images might lead to unnatural results as shown in Fig.~\ref{fig:curicture_texture}. To solve this problem, we formulate the texture fusion from multi-view caricatures as a label problem.
%\begin{figure}[t]
%\begin{center}
%\includegraphics[width=1.0\linewidth]{pdfs/tex.pdf}
%\end{center}
%   \caption{Given a set of caricatures from different views, we first select pixels by a graph cut based optimization method and generate a complete caricature texture map using poisson integration.}
%\label{fig:fusion_pipeline}
%\end{figure}

\begin{figure}[t]
\begin{center}
\includegraphics[width=1.0\linewidth]{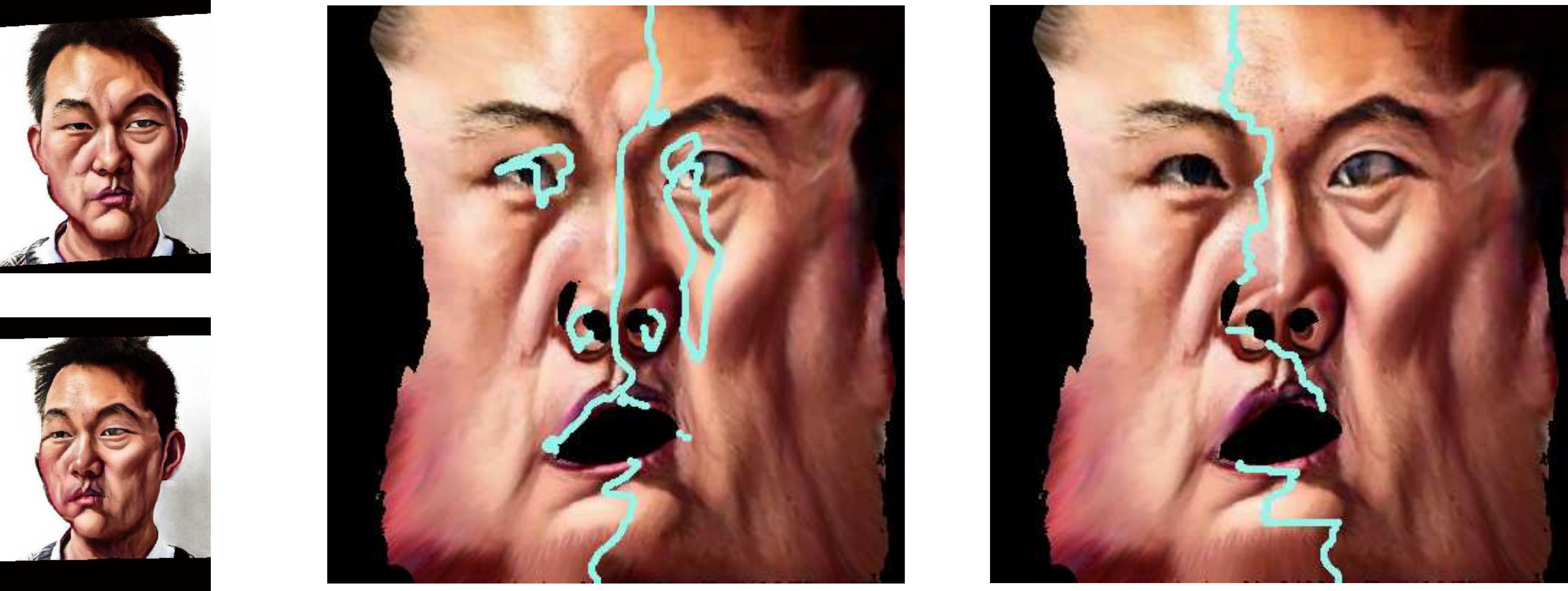}
\end{center}
   \caption{Given two caricatures from different views in the first column, we fuse them together to generate a texture map. The second column shows the fusion result only considering the term in Eq.~\eqref{eq:viewsimilarity}, and the third column shows the fusion result with our graph cut based optimization method by optimizing Eq.~\eqref{eq:mathcingcost} and Eq.~\eqref{eq:viewsimilarity}. The cyan lines are the labeling boundaries.}
\label{fig:compare_cut}
\end{figure}

Similar with the approach adopted in~\cite{CasasVCH14,DuCCHV18}, which formulate the image stiching problem as a labeling problem and solve it with graph cut, we generate a complete texture map from multi-view caricature images based on a labeling formulation. We first define a matching cost for pixels from two images for the fusion problem. For the images $\bfI_i$ and $\bfI_j$ which have overlaps, let $\bfu_1$ and $\bfu_2$ be two adjacent pixels in the overlap region, and we define the matching cost of the pixel pair as:
\begin{equation}
\label{eq:mathcingcost}
\bfD(\bfu_1, \bfu_2, \bfI_{i}, \bfI_{j}) = \|\bfI_{i}(\bfu_1) - \bfI_{j}(\bfu_1)\|_2 + \|\bfI_{i}(\bfu_2) - \bfI_{j}(\bfu_2)\|_2,
\end{equation}
where $\bfI(\bfu)$ is the normalized pixel color at position $\bfu$. The above term could make the fusion between different views to be smooth, while avoiding the fusion line passing through feature areas like eye, nose and mouth. On the other hand, the fusion should also consider the views of images, and thus we define the following data term:
\begin{equation}
\label{eq:viewsimilarity}
\bfD(\bfu, \bfI_i) = 2 - \|\bfn_i(\bfu) - \bfd_{i}\|_{2},\quad \bfD(\bfu, \bfI_j) = 2 - \|\bfn_j(\bfu) - \bfd_{j}\|_{2},
\end{equation}
where $\bfn_i(\bfu)$ is the vertex normal corresponding to pixel $\bfu$ in mesh reconstructed from image $\bfI_i$, and $\bfd_{i}$ and $\bfd_{j}$ are the view directions of images $\bfI_i$ and $\bfI_j$ respectively optimized by solving Eq.~\eqref{eq:bundle}. The term in Eq.~\eqref{eq:viewsimilarity} is used to measure the angle between surface normals and the view ray.

\begin{figure}[t]
\begin{center}
\includegraphics[width=1.0\linewidth]{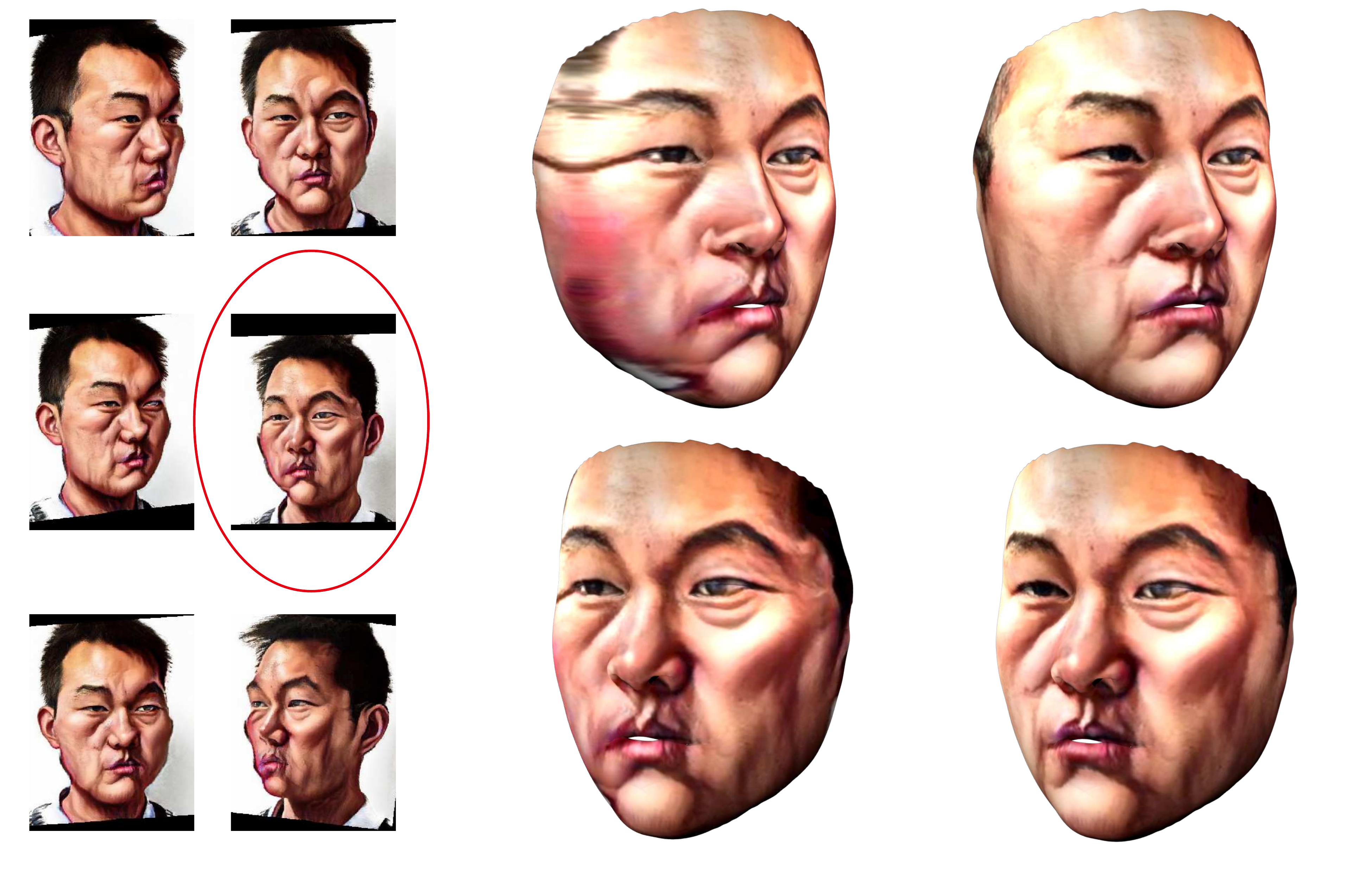}
\end{center}
   \caption{The first two columns show input caricatures, the third column shows the rendered models using texture generated by one caricature(the circled one), and the last column shows the rendered models using texture generated by multiview caricatures. With multiview caricatures, we can generate a more complete texture.}
\label{fig:one_view}
\end{figure}

Similar with the MRF optimization problem in~\cite{CasasVCH14,DuCCHV18}, we apply graph cut~\cite{BK:pami04} to solve this labeling problem with smoothness term in Eq.~\eqref{eq:mathcingcost} and data term in Eq.~\eqref{eq:viewsimilarity}, and we multiply a weight 1.2 to Eq.~\eqref{eq:viewsimilarity} to balance the importance of these two terms.  Fig.~\ref{fig:compare_cut} shows the comparison result between with and without Eq.~\eqref{eq:mathcingcost} with the same inputs. We can observe that the fusion result with both Eq.~\eqref{eq:mathcingcost} and Eq.~\eqref{eq:viewsimilarity} is smooth and the features like nose are preserved. Finally, in order to generate a complete caricature texture map, we integrate the selected texture from different views into the template texture map using poisson integration~\cite{perez2003poisson}. In Fig.~\ref{fig:curicture_texture}, we compare the complete caricature texture map results by these two methods, and we can observe that the results by our proposed method are more satisfying especially on the areas marked by the red box. We also compare our method with using only one caricature to generate complete texture in Fig.~\ref{fig:one_view}, and it is obvious that one caricature based approach can not generate satisfying result.

\section{Dynamic Modeling}
\label{sec:dynamic}

During tracking process, we first reconstruct the 3D face shape for each frame, and then translate the 3D shape from regular style to caricature style.

\subsection{Face Tracking}
In static modeling, we already obtained the blendshape $\bfB$ for the user. A new face expression is generated as $\bfF(\bfB, \bfw) = \bfb_0 + \Delta\bfB\bfw$, where $\Delta\bfB = [\bfb_1 - \bfb_0, \ldots , \bfb_n - \bfb_0]$ and $\bfw=[w_1, \ldots ,w_n]^{T}$ are blendshape weights. For each frame, we can recover its 3D face shape by optimizing its blendshape weights $\bfw$.

We employ the three energy terms in~\cite{ichim2015dynamic} to estimate the rigid face pose $(\bfR, \bft)$ and blendshape weights $\bfw^k$ at $k$-th frame. The first term is the facial feature energy which is formulated as
\begin{equation}
	E_{\textrm{fea}} = \sum_{i=1}^{L} \|\Pi(\bfR\bfF_{v_i}(\bfB, \bfw^{k}) + \bft) - \bfu_{i}\|_2^2,
	\label{feature_alignment}
\end{equation}
where $\bfF_{v_i} \in \mathbb{R}^{3}$ and $\bfu_{i} \in \mathbb{R}^{2}$ are the coordinates of the $i$-th 3D landmark vertex and the corresponding image landmark, $\Pi(\cdot)$ projects a 3D point to a 2D point. The second term is the texture-to-frame optical flow energy that can improve the robustness of lighting variations, and defined as 
\begin{equation}
	E_{\textrm{flow}} = \sum_{i, j} \left\| 
	\left[
	\begin{matrix}
	\rho_{i, j+1} - \rho_{i, j}\\
	\rho_{i+1, j} - \rho_{i, j}\\
	\end{matrix}
	\right] \!\!-\!\!
	\left[
	\begin{matrix}
	I(\bfu_{v_{i, j+1}}) - I(\bfu_{v_{i, j}})\\
	I(\bfu_{v_{i+1, j}}) - I(\bfu_{v_{i, j}})\\
	\end{matrix} 
	\right]
	\right\|_2^2,
	\label{eq:optical_flow}
\end{equation}
where $\bfu_{v} = \Pi(\bfR\bfF_{v}(\bfB, \bfw^k) + \bft)$, and $\{v_{i, j}\}$ is a set of visible points located on the mesh surface at $k$-th frame. $\rho_{i, j}$ is the  gray-scale value at location $(i,j)$ in the albedo texture obtained in the Sec.~\ref{sssec:blendshape}, and $I(\bfu)$ is the gray-scale color at location $\bfu$ in current frame. The position $(i, j)$ is a 2D parametric coordinate of the mesh vertex $v_{i,j}$. The third term is $\ell_{1}$-norm regularization on the blendshape coefficients
\begin{equation}
	E_{\textrm{spa}} = \|\bfw^k\|_1.
	\label{eq:sparse}
\end{equation}
This is because the blendshape basis are not linearly independent and this sparsity-inducing energy can avoid potential blendshape compensation artifacts to stabilize the tracking. Except for the sparsity regularization, we also smooth the face shape in the temporal domain by
\begin{equation}
	E_{\textrm{sm}} = \|\bfw^{k-2} - 2\bfw^{k-1} + \bfw^{k}\|_2^2.
\end{equation}
Finally, our facial tracking energy can be formulated as 
\begin{equation}
	E_{\textrm{tracking}} = E_{\textrm{fea}} + \mu_{\textrm{flow}}E_{\textrm{flow}} + \mu_{\textrm{spa}}E_{\textrm{spa}} + \mu_{\textrm{sm}}E_{\textrm{sm}},
	\label{tracking}
\end{equation}
where $\mu_{\textrm{flow}}, \mu_{\textrm{spa}}, \mu_{\textrm{sm}}$ are user-specified weights. To optimize this problem, we first set $\bfw$ to zero and optimize the rigid pose ($\bfR, \bft$). Then we use a warm started shooting method~\cite{fu1998penalized} to optimize blendshape weights $\bfw$ while fixing ($\bfR, \bft$). This process is iterated three times in our implementation.

\subsection{3D Style Translation}
One might think that translating a 3D face expression sequence from regular style to caricature style is an easy task, such as directly applying the method proposed in~\cite{GaoYQLRXX18}. However, different from the examples used in~\cite{GaoYQLRXX18} whose variance is mainly about expressions or actions, 3D caricature models not only have different identities, expressions but also different styles. Besides, we require that the translated 3D caricature model has similar expression with the original 3D face shape for each frame, and the identities for all the generated models in the sequence should be same, otherwise, the generated 3D caricature sequence is not visually natural.

\begin{figure}[t]
\begin{center}
\includegraphics[width=1.0\linewidth]{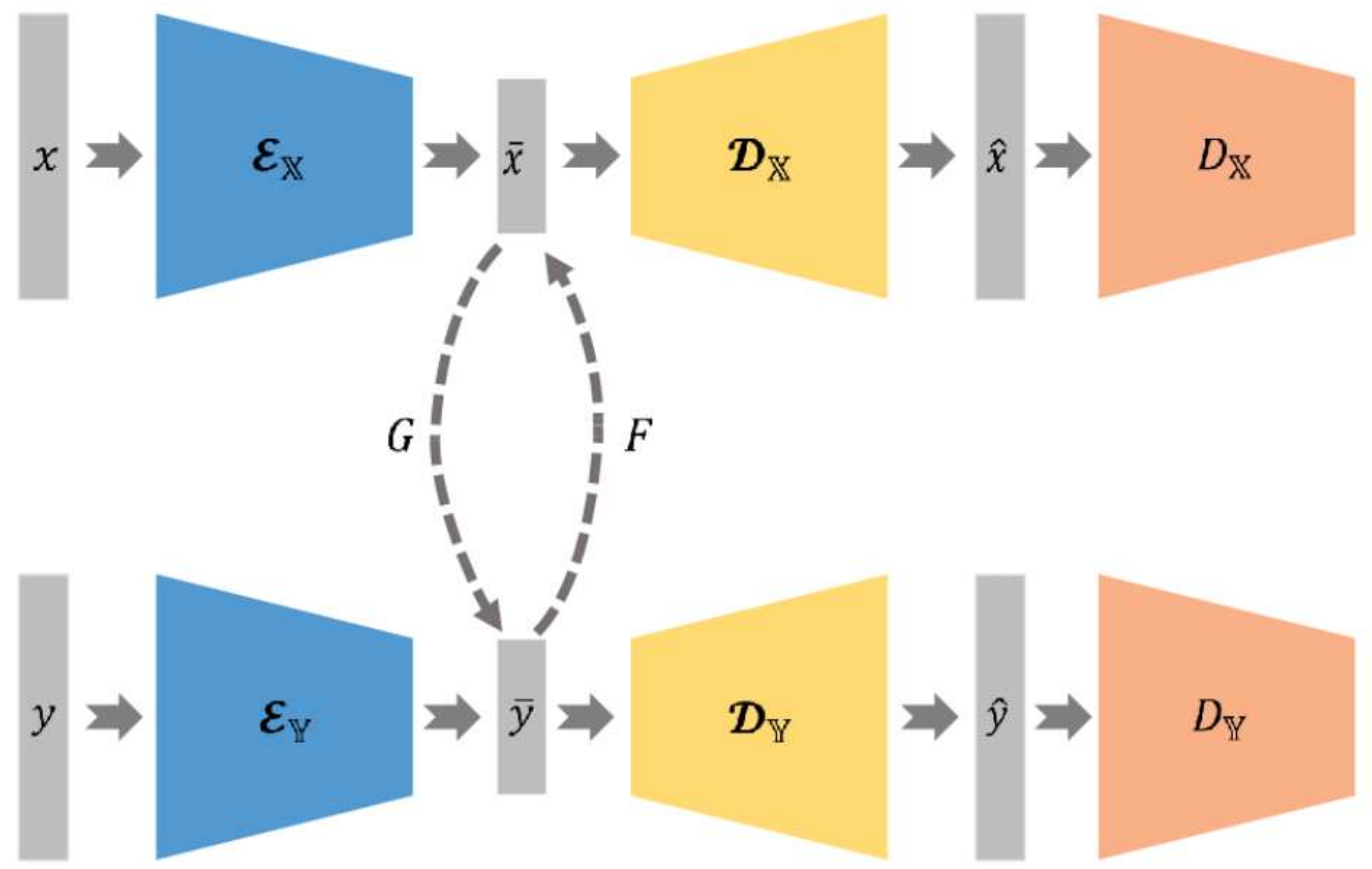}
\end{center}
   \caption{The overall architecture of our VAE Cycle-Consistent Adversarial (VAE-CycleGAN)
Network for 3D style transfer.}
\label{fig:VAE-Structure}
\end{figure}

To solve this challenging problem, two well-designed variational autoencoders (VAE) and one CycleGAN model are trained to translate the 3D face shapes from regular domain $\bbX$ to caricature domain $\bbY$ in each frame. We first train VAE models to encode regular shapes and caricature shapes into compact latent spaces $\overline{\bbX}$ and $\overline{\bbY}$, and then train CycleGAN model to learn two mapping functions between regular shapes and caricature shapes in the latent spaces ($\overline{\bbX} \leftrightarrows \overline{\bbY}$). It is easier and more reliable to learn the translation between regular style and caricature style in latent spaces than in original shape spaces as the VAE constructs an embedding space of 3D face shapes. Fig.~\ref{fig:VAE-Structure} shows the overall network architecture.

\begin{figure*}[t]
\begin{center}
\includegraphics[width=1\linewidth]{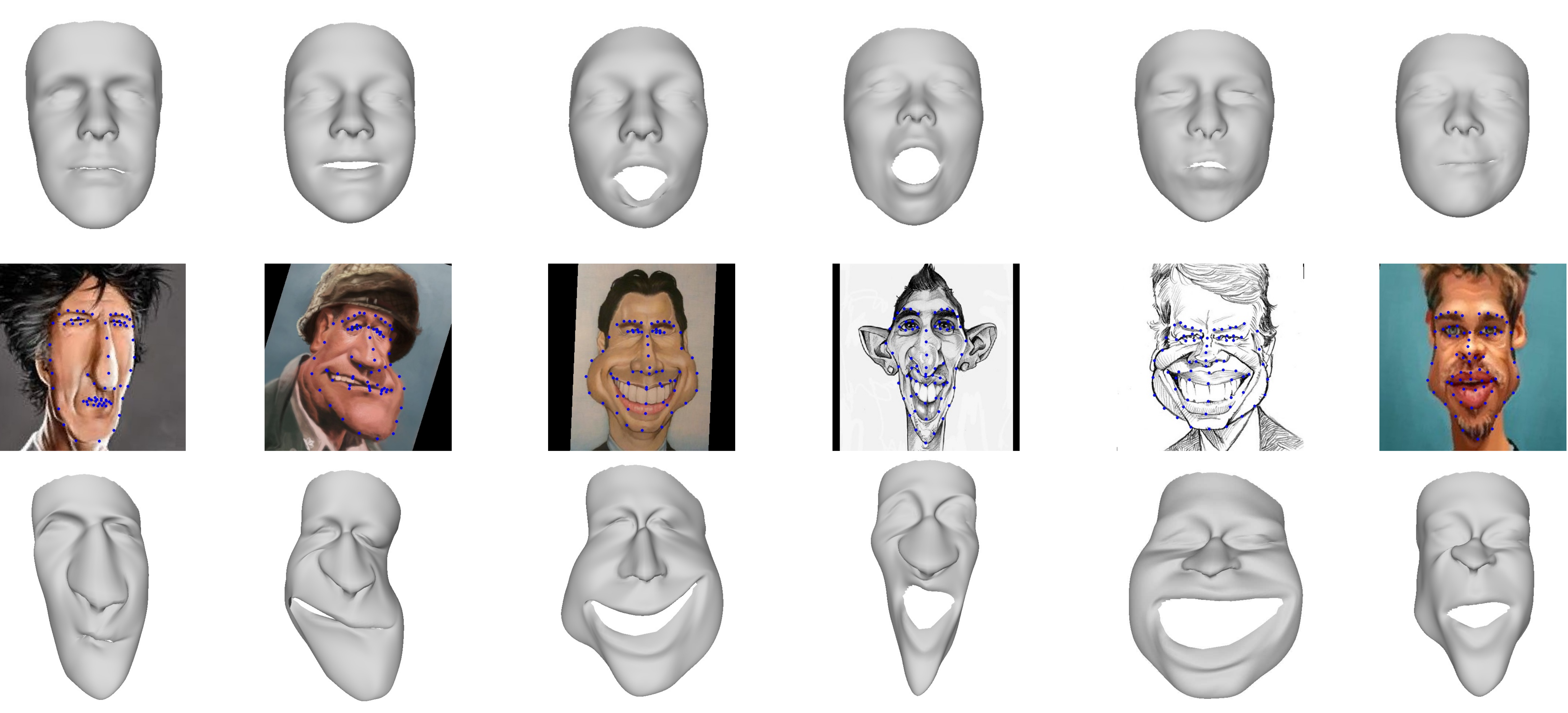}
\end{center}
\caption{Samples of our training data. From top to bottom: samples in regular domain, samples of collected portrait caricatures with manually labeled $68$ landmarks, corresponding 3D exaggerated shapes using the method in ~\cite{WuZLZC18}.}
\label{fig:training_samples}
\end{figure*}

\begin{figure*}[t]
\begin{center}
\includegraphics[width=1\linewidth]{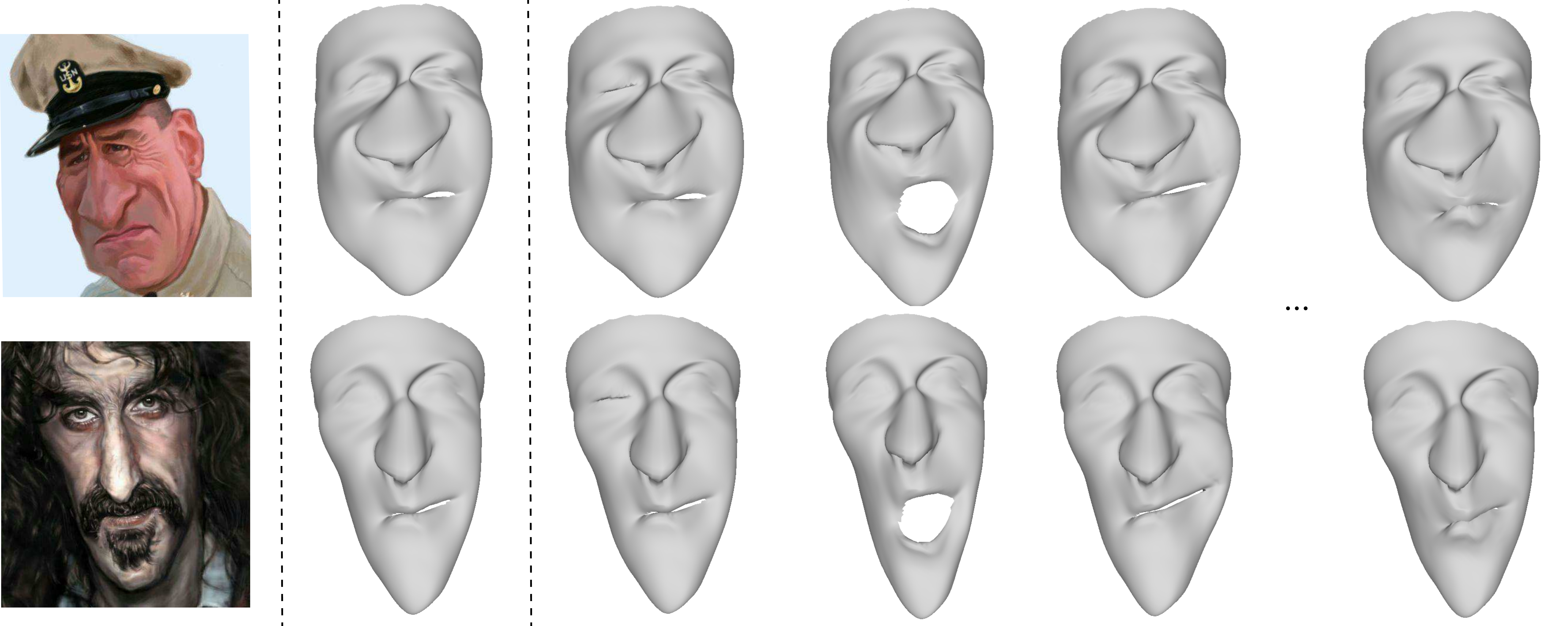}
\end{center}
\caption{Samples of our augmented caricature training data. From left to right: the input 2D caricatures, the reconstructed 3D caricature models and augmented models with different expressions.}
\label{fig:augmentdata}
\end{figure*}

\subsubsection{3D Shape Training Set} Before training VAEs in regular domain $\bbX$ and caricature domain $\bbY$ respectively, we first construct two sets of unpaired face shapes as our training data. For the regular face domain $\bbX$, we collect photos with neutral expression from different views and an expressive video for each of 30 people. Through our static and dynamic face modeling, we obtain a set of face models with continuous expression for each people. 100 expressive face models are selected at intervals from each set, for a total of 3,000 face models. To further enrich the identities of training data, we register a template face model non-rigidly with 150 neutral face shapes from FaceWareHouse database by using Laplacian deformation~\cite{sorkine2004laplacian}, and then apply deformation transfer~\cite{sumner2004deformation} from the above selected expressions to the new 150 face models. 50 selected expressions are used to construct 7,500 new face models and thus our training data set in total includes 10,500 face models at last. For the caricature domain $\bbY$, we use the method in~\cite{WuZLZC18} to reconstruct 3D exaggerated models on our collected 6,000 portrait caricatures in Sec.~\ref{2d_cari_gen}. Training shapes in regular domain and caricature domain have the same connectivity with $11865$ vertices and $23250$ triangles. Some training samples are shown in Fig.~\ref{fig:training_samples}.

\noindent{\textbf{Data Augmentation.}} Although our constructed caricature database contains rich identities and caricature styles, the expressions of caricatures are relatively simple, mainly about neutral expression and laughing mouths. However, in our real face video data, it contains micro-expressions such as pouting and frowning. In order to enable our trained translation network to translate expressions from regular style to caricature style more naturally, we propose a data augmentation strategy to enhance the exaggerated 3D face models. First, we manually select 1,000 neutral face models with different identities from the constructed caricature database. And then we apply deformation transfer~\cite{sumner2004deformation} from the FaceWareHouse~\cite{cao2014facewarehouse} expressions to these neutral 3D caricatures. Thus we construct a new expressive caricature database of 47,000 face models. Some samples of the augmented models are shown in Fig.~\ref{fig:augmentdata}.

In summary, we construct a training database which includes 10,500 face models of 180 identities for the regular domain $\bbX$. We construct two training databases for the caricature domain $\bbY$, the first one includes 6,000 exaggerated face models, and the second one includes 47,000 exaggerated face models of 1,000 identities.

\subsubsection{VAE-embedding} 
\begin{figure*}[t]
\begin{center}
\includegraphics[width=1\linewidth]{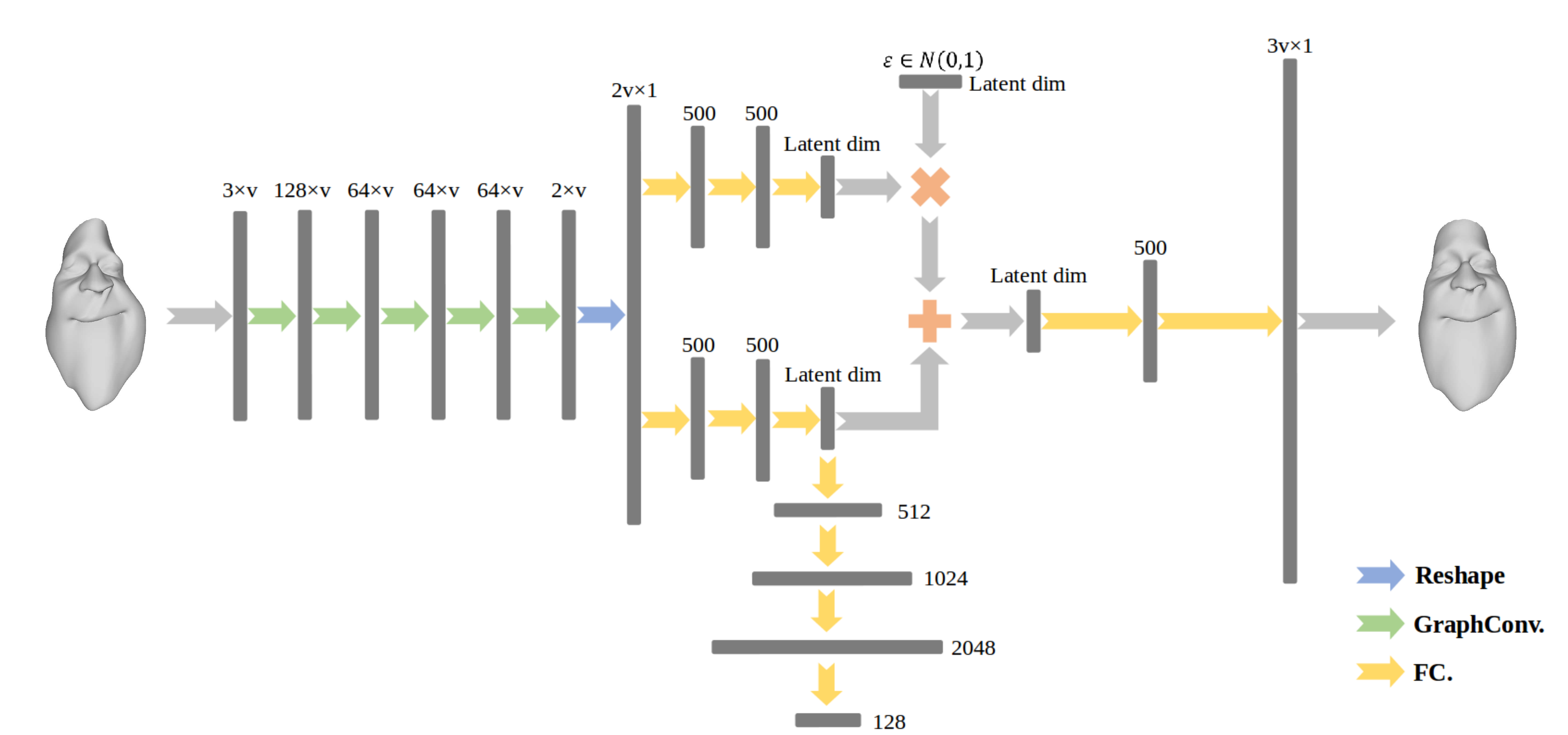}
\end{center}
\caption{Structure of our VAE network. The latent dim is 200 and 350 for regular domain and caricature domain respectively. Besides original VAE structure, the latent code is also connected with some fully connected layers to regress a $128$-dim feature for face recognition.}
\label{fig:vae_net}
\end{figure*}

After constructing these training sets of unpaired face models, we train a VAE for each domain. Our VAE network consists of an encoder $\calE(\cdot)$ and a decoder $\calD(\cdot)$. Given a face shape $\bfx \in \bbX$, $\overline{\bfx} = \calE_{\bbX}(\bfx)$ is the encoder latent vector and $\hat{\bfx} = \calD_{\bbX}(\overline{\bfx})$ is reconstructed face shape. $\calE_{\bbY}(\cdot)$ and $\calD_{\bbY}(\cdot)$ are the encoder and decoder networks respectively in the caricature domain $\bbY$. Different from~\cite{GaoYQLRXX18}, we directly use vertex coordinates of the mesh rather than ACAP shape deformation representation~\cite{tan2017mesh} as ACAP feature extraction is time-consuming and thus can not achieve real-time. We adopt the graph convolutional layer~\cite{defferrard2016convolutional} in our VAE network, and the network structure is shown in Fig.~\ref{fig:vae_net}. Our well-designed VAE network includes the following loss terms.

\noindent{\textbf{Reconstruction Loss.}} The first loss is the MSE (mean square error) reconstruction loss that requires the reconstructed face shape to be same as the original face shape
\begin{equation}
	L_{\textrm{rec}} = \frac{1}{|\bbX|}\sum_{\bfx\in\bbX}\|\hat{\bfx}-\bfx\|_1.
\end{equation}

\noindent{\textbf{Regularization Loss.}} The second loss is the KL divergence to promote Gaussian distribution in the latent space
\begin{equation}
	L_{\textrm{KL}} = D_{\textrm{KL}}(q(\overline{\bfx}|\bfx)|p(\bfx)),
\end{equation}
where $q(\overline{\bfx}|\bfx)$ is the posterior distribution given input shape and $p(\bfx)$ is the Gaussian prior distribution.

\begin{figure}[t]
\begin{center}
\includegraphics[width=1.0\linewidth]{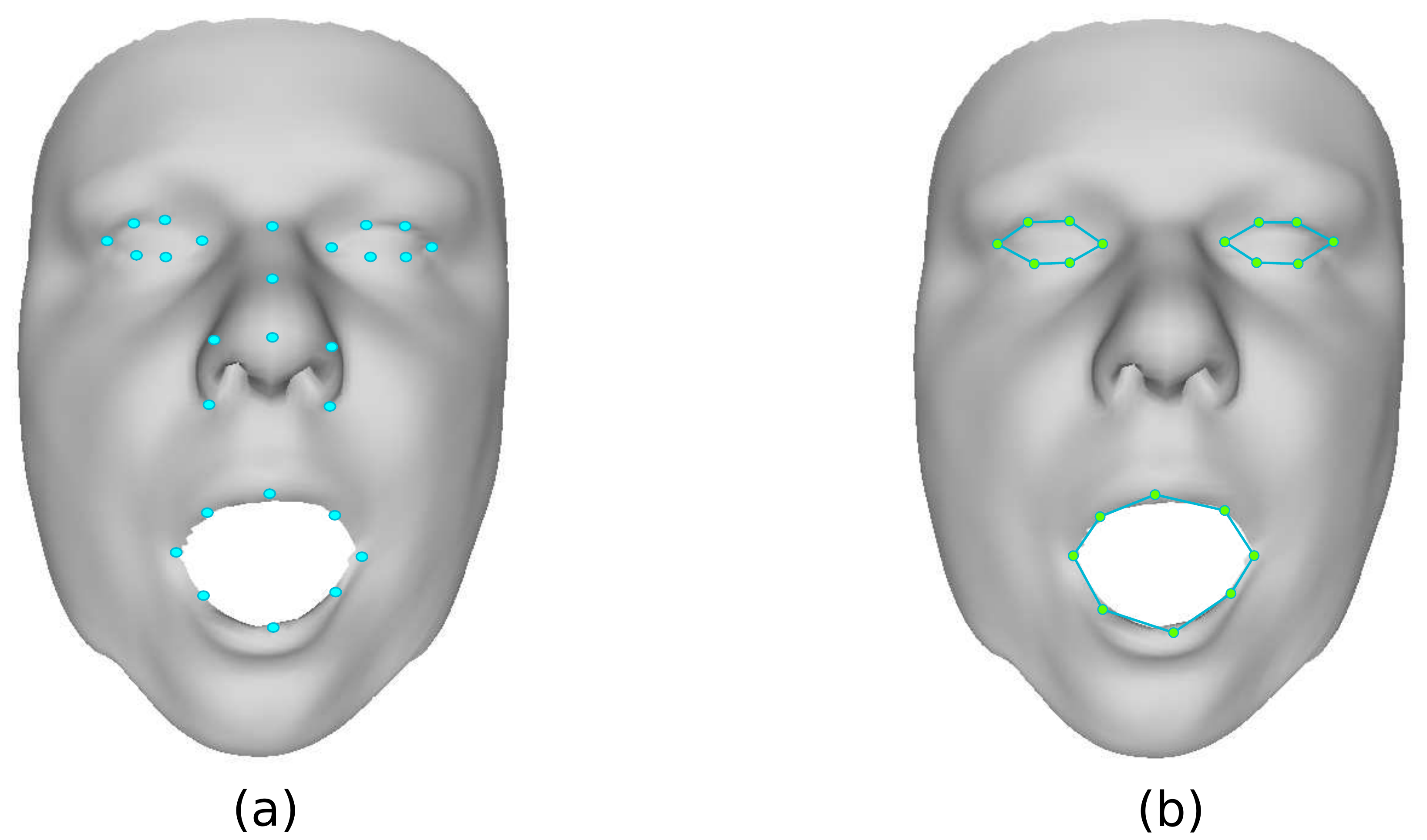}
\end{center}
\caption{(a) shows the landmark vertices used for keeping the expressions of the reconstructed shapes and original shapes are the same. (b) shows angles used to enforce that the expression of the reconstructed result $\calD_{\bbY}(G(\overline{\bfx}))$ should be similar to that of the real sample $\bfx$ in domain $\bbX$. We use all angles of the three polygons.}
\label{fig:landmarks}
\end{figure}

\noindent{\textbf{Expression Loss.}} With only the above two losses which are often used in VAE networks, it can not guarantee that expressions of the reconstructed shape and original shape are the same. Therefore we add a new feature alignment loss to constrain their expressions
\begin{equation}
	L_{\textrm{exp}} = \frac{1}{L}\sum_{i=1}^{L}\|\hat{\bfx}(i) - \bfx(i)\|_2^2,
\end{equation}
where $\{\bfx(i)\}_{i=1}^{L}$ and $\{\hat{\bfx(i)}\}_{i=1}^{L}$ are the landmark vertices on the original shape and reconstructed shape respectively. The landmark vertices are shown in Fig~.\ref{fig:landmarks} (a). Thus the VAE loss can be formulated as
\begin{equation}
	L_{\textrm{VAE}} = L_{\textrm{rec}} + \mu_{\textrm{KL}}L_{\textrm{KL}} + \mu_{\textrm{exp}}L_{\textrm{exp}},
\end{equation}
where $\mu_{\textrm{KL}}$ and $\mu_{\textrm{exp}}$ are user-specified weight parameters. This VAE loss is similarly defined for the caricature domain $\bbY$. Note that we train the VAE network on both of the two caricature training databases for domain $\bbY$. 

After training the VAE models, the encoder latent vectors $\calE_{\bbX}(\bfx)$ and $\calE_{\bbY}(\bfy)$ are used to train the CycleGAN network. 
Thus we can use the VAE models and CycleGAN model to translate a regular face model $\bfx \in \bbX$ to a reconstructed caricature $\hat{\bfy}\in \bbY$. Let $\bfx_1$ and $\bfx_2$ denote two different expressions of the same identity respectively, their corresponding reconstructed caricatures $\hat{\bfy}_1$ and $\hat{\bfy}_2$ may have different identities.

\begin{figure}
\begin{center}
\includegraphics[width=1.0\linewidth]{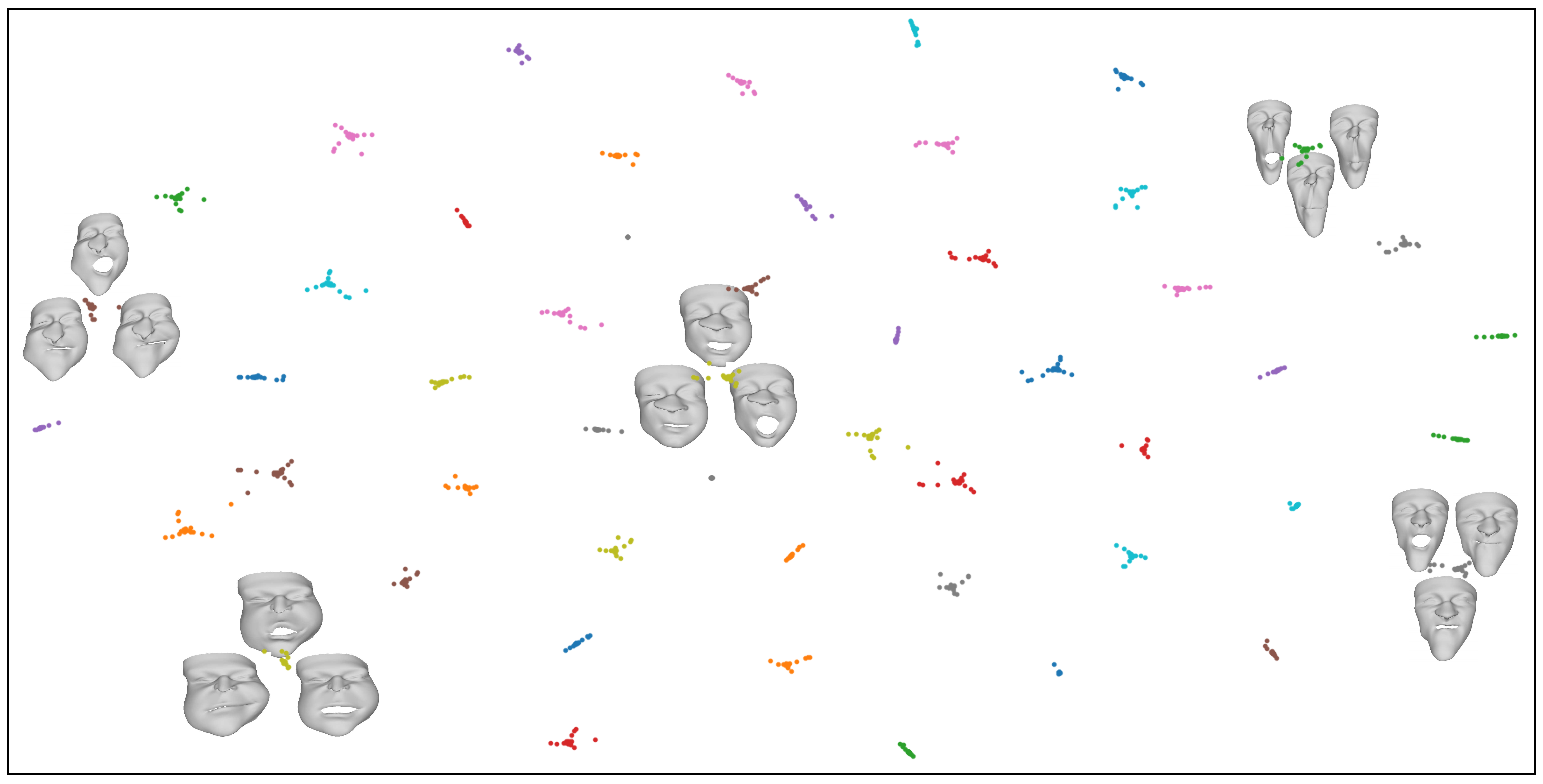}
\end{center}
\caption{T-SNE visualization of the embedding space of caricature shapes. Here we show the latent codes of $50$ identities with different colors, and we show $5$ identities with different expressions around their corresponding codes.}
\label{fig:vae_space}
\end{figure}

\noindent{\textbf{Identity Loss.}}
In order to preserve the identity of reconstructed caricatures from different expressions of the same person, we first constrain that the latent codes of different expressions of the same identity should be close in each domain. For this purpose, we add a 3D face recognition loss to help train the VAE-embedding, which enforces separation of features from different classes and aggregation of features within the same class. Our face recognition network consists of two parts, the first part is the encoder network in VAE, the second part is a multi-layer neural network which stacks several fully connected layers, see Fig.~\ref{fig:vae_net}. The latent codes are fed into this multi-layer neural network. We denote the output of our face recognition network as $f(\bfx_i)$ and the $i$-th face model $\bfx_i$ corresponding label by $l_i$. Different expressions of the same person have the same label. The face recognition loss is defined as
\begin{equation}
\footnotesize
	\begin{aligned}
		L_{\textrm{id}}=-\sum_{i=1}^{S}\log(\frac{\exp(\bmalpha_{l_i}^{T}f(\bfx_i)+\beta_{l_i})}{\sum_{j=1}^{I}\exp(\bmalpha_{j}^{T}f(\bfx_i)+\beta_j)})+ \frac{\mu_c}{2}\|f(\bfx_i)-\bfc_{l_i}\|_2^2,
	\end{aligned}
\end{equation}
where $\bmalpha=[\bmalpha_1,\ldots,\bmalpha_I]\in\mathbb{R}^{K\times I}$ and $\bmbeta=[\beta_1,\ldots,\beta_I]\in \mathbb{R}^{1\times I}$ are the weights and biases in the last fully connected layer. $\bfC=[\bfc_1,\ldots, \bfc_I] \in \mathbb{R}^{K\times I}$ are the weights in the center loss~\cite{wen2016discriminative} representing the feature center for each identity. $S$ is the batch size, $K$ is the dimension of the output feature of our face recognition network, $I$ is the number of identities and $\mu_c$ is the balance weight.

At last, we jointly train the encoder and decoder in the VAE network by optimizing the new VAE loss
\begin{equation}
	L_{\textrm{VAE}}^{*} = L_{\textrm{VAE}} + \mu_\textrm{id}L_{\textrm{id}},
\end{equation}
where $\mu_{\textrm{id}}$ is a user-specified weight. This new VAE loss is similarly defined for the caricature domain $\bbY$. Note that for the domain $\bbY$, our first caricature training data is feed into the $L_{\textrm{VAE}}$ and our second caricature training data is feed into the $L_{\textrm{id}}$. We show T-SNE visualization of the embedding space of caricature shapes in Fig.~\ref{fig:vae_space}. It is shown that with the well designed identity loss $L_{\textrm{id}}$, the latent codes of different expressions of the same identity are closer than that of different identities.

\subsubsection{Style Translation} Now we train two mapping functions $G: \overline{\bbX} \rightarrow \overline{\bbY}$ and $F: \overline{\bbY} \rightarrow \overline{\bbX}$ by using the CycleGAN network with four loss terms. The first is adversarial loss $L_{\textrm{adv}}(G, D_{\bbY}, \overline{\bbX}, \overline{\bbY})$, which makes the reconstructed results $\calD_{\bbY}(G(\overline{\bfx}))$ from the translated codes $G(\overline{\bfx})$ identical to the real sample in domain $\bbY$
\begin{equation}
	\begin{aligned}
	L_{\textrm{adv}}(G, D_{\bbY}, \overline{\bbX}, \overline{\bbY}) &= \mathbb{E}_{\hat{\bfy}\sim P_{\textrm{data}}(\hat{\bbY})}[log(D_{\bbY}(\hat{\bfy}))]\\ &+ \mathbb{E}_{\overline{\bfx}\sim P_{\textrm{data}}(\overline{\bbX})}[log(1-D_{\bbY} (\calD_{\bbY}(G(\overline{\bfx})))],
	\end{aligned}
\end{equation}
where $D_{\bbY}$ is a discriminator to distinguish the generated samples $\calD_{\bbY}(G(\overline{\bfx}))$ from the real ones in $\bbY$. $P_{\textrm{data}}(\cdot)$ represents the data distribution and $\mathbb{E}$ is the expected value of the distribution. The adversarial loss is similarly defined and denoted by $L_{\textrm{adv}}(F, D_{\bbX}, \overline{\bbX}, \overline{\bbY})$.

The second loss is cycle-consistency loss, which makes the latent codes translation cycle to be able to bring $\bfx$ back to the original code, i.e. $\overline{\bfx} \rightarrow G(\overline{\bfx}) \rightarrow F(G(\overline{\bfx})) \approx \overline{\bfx}$. The cycle-consistency loss for the mapping $\bbX \rightarrow \bbY$ is formulated as
\begin{equation}
	L_{\textrm{cyc}}(G, F) = \mathbb{E}_{\overline{\bfx}\sim P_{\textrm{data}}(\overline{\bbX})}[\|F(G(\overline{\bfx}))-\bfx\|_1].
\end{equation}
The cycle-consistency loss for the inverse mapping $\bbX \rightarrow \bbY$ can be defined in a similar way and denoted by $L_{\textrm{cyc}}(F, G)$.

The third loss is feature alignment loss, which enforces that the expression of the reconstructed result $\calD_{\bbY}(G(\overline{\bfx}))$ should be similar to that of the real sample $\bfx$ in domain $\bbX$. We use a vector of angles between 3D adjacent landmarks to describe the expression of a face model, see Fig.~\ref{fig:landmarks} (b). Let $\Theta(\cdot)$ denote this vector, where $\cdot$ denotes a face model in domain $\bbX$ or $\bbY$. For the mapping $G$, our feature alignment loss is defined as
\begin{equation}
	L_{\textrm{ang}}(G, \Theta) = \|\Theta(\calD_{\bbY}(G(\overline{\bfx}))) - \Theta(\bfx)\|_2^2.
\end{equation}
$L_{\textrm{ang}}(F, \Theta)$ is defined symmetrically.

The fourth loss is contrastive loss~\cite{hadsell2006pair}, which constrains that a pair of translated codes $G(\overline{\bfx}_i), G(\overline{\bfx}_j)$ should be close if they belong to the same identity, and they should be far away otherwise. It is formulated as 
\begin{equation}
	\begin{aligned}
	L_{\textrm{pair}}(G) &= \sum_{i,j}  \lambda L_{\textrm{dist}}(G(\overline{\bfx}_i), G(\overline{\bfx}_j))\\ &+ (1-\lambda)\max(\tau-L_{\textrm{dist}}(G(\overline{\bfx}_i), G(\overline{\bfx}_j)), 0),
	\end{aligned}
\end{equation}
where $L_{\textrm{dist}}(G(\overline{\bfx}_i), G(\overline{\bfx}_j)) = 1-\frac{G(\overline{\bfx}_i)\cdot G(\overline{\bfx}_j)}{\|G(\overline{\bfx}_i)\|_2\|G(\overline{\bfx}_j)\|_2}$ is the cosine distance, $\lambda=1$ if $l_i=l_j$, vice versa, and $\tau$ is a user-specified margin. $L_{\textrm{pair}}(F)$ is defined similarly. We construct some pairs of translated codes from the current batch. Note that we train this contrastive loss only on the second caricature training database. 

Finally, we train the CycleGAN network by optimizing the joint loss
\begin{equation}
	\begin{aligned}
	L_{\textrm{CycleGAN}} &= (L_{\textrm{adv}}(G, D_{\bbY}, \overline{\bbX}, \overline{\bbY}) + L_{\textrm{adv}}(F, D_{\bbX}, \overline{\bbX}, \overline{\bbY}))\\ &+ \mu_{\textrm{cyc}}(L_{\textrm{cyc}}(G, F) + L_{\textrm{cyc}}(F, G))\\ &+ \mu_{\textrm{ang}}(L_{\textrm{ang}}(G, \Theta) + L_{\textrm{ang}}(F, \Theta))\\ &+ \mu_{\textrm{pair}}(L_{\textrm{pair}}(G)+L_{\textrm{pair}}(F)),
	\end{aligned}
\end{equation}
where $\mu_{\textrm{cyc}}$, $\mu_{\textrm{ang}}$ and $\mu_{\textrm{pair}}$ are the user-specified weight parameters. 

\subsubsection{Temporal Smoothing} After training the VAE-CycleGAN, we can translate the 3D face shape from $X$ to $Y$ in each frame. To further preserve the temporal smoothness of the identity and expression of the user, we add a smooth regularization in the latent space $\overline{\bbY}$ rather than the 3D face shape space. Given a regular face model $\bfx^k$ in the current frame $k$, we can obtain its latent code translation $G(\calE_{\bbX}(\bfx^k))$. Now a new latent code $\overline{\bfx}^{*}$ can be solved by optimizing the following problem about variable $\overline{\bfx}^{*}$
\begin{equation}
\footnotesize
	\begin{aligned}
	\mu_{smo}\|G(\calE_{\bbX}(\bfx^{k-2})) -2G(\calE_{\bbX}(\bfx^{k-1})) + \overline{\bfx}^{*}\|_2^2 + \|\overline{\bfx}^{*} - G(\calE_{\bbX}(\bfx^k))\|_2^2,
	\end{aligned}
\end{equation}
where $\mu_{smo}$ is a balance weight. The code $\overline{\bfx}^{*}$ is then used to decoder a 3D caricature $\calD_{\bbY}(\overline{\bfx}^{*})$ at the $k$-th frame.
\section{Experimental Results}
\label{sec:results}

\subsection{Implementation Details}
In the experiments, we fix the weight parameters $\mu_{\textrm{flow}} = 1$, $\mu_{\textrm{spa}} = 10$, $\mu_{\textrm{sm}} = 0.001$, $\mu_{\textrm{KL}} = 1e^{-5}$, $\mu_{\textrm{exp}} = 5$,  $\mu_{\textrm{id}} = 1$, $\mu_{\textrm{cyc}} = 10$, $\mu_{\textrm{ang}} = 5$, $\mu_{\textrm{pair}} = 1$, $\mu_{\textrm{smo}} = 0.001$. We train all our networks with PyTorch~\cite{paszke2017automatic} framework. The VAEs and CycleGAN are trained with 200 epochs using Adam solver~\cite{kingma2014adam} and we set batch size to 50 and base learning rate to 0.0001. Our static modeling, real-time dynamic modeling and 3D style translation are conducted on a PC with Intel i7-4790 CPU, 8GB RAM and NVIDIA GTX 1070 GPU. It takes about 3 minutes to do static modeling for a user with 5 $1024 \times 1024$ images, and the outputs include the 3D face blendshape for the user and the complete caricature texture map. For each frame with size $512 \times 512$, it takes less than 10ms for 3D face modeling and about 40ms for 3D style translation.

\begin{figure}[t]
\begin{center}
\includegraphics[width=0.9\linewidth]{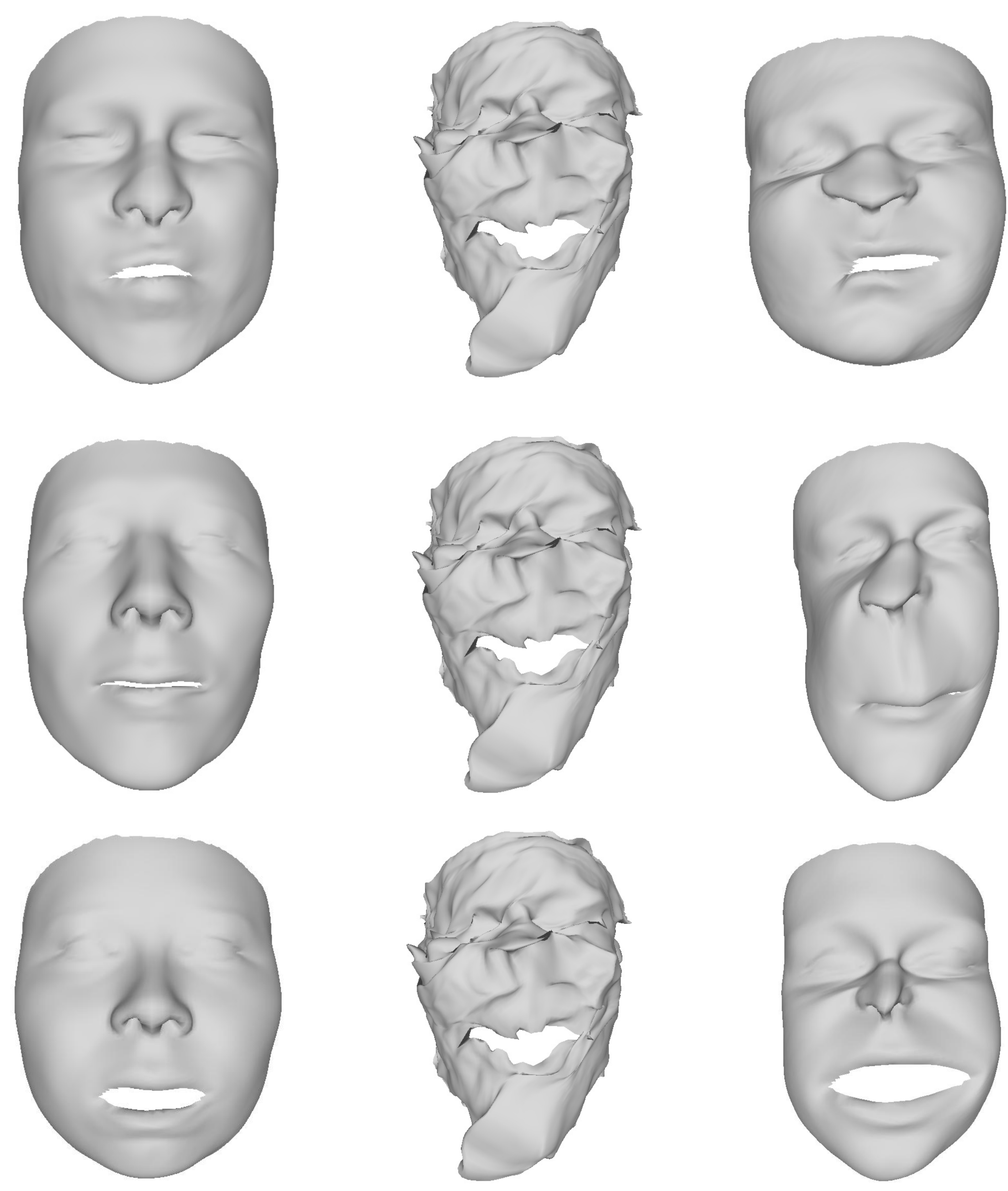}
\end{center}
\caption{Comparison of CycleGAN learned in latent spaces and original shape spaces. From left to right: input regular 3D face, transferred result of CycleGAN learned in original shape spaces, and transferred result of CycleGAN learned in latent spaces.}
\label{fig:novae}
\end{figure}

\begin{figure}[t]
\begin{center}
\includegraphics[width=0.8\linewidth]{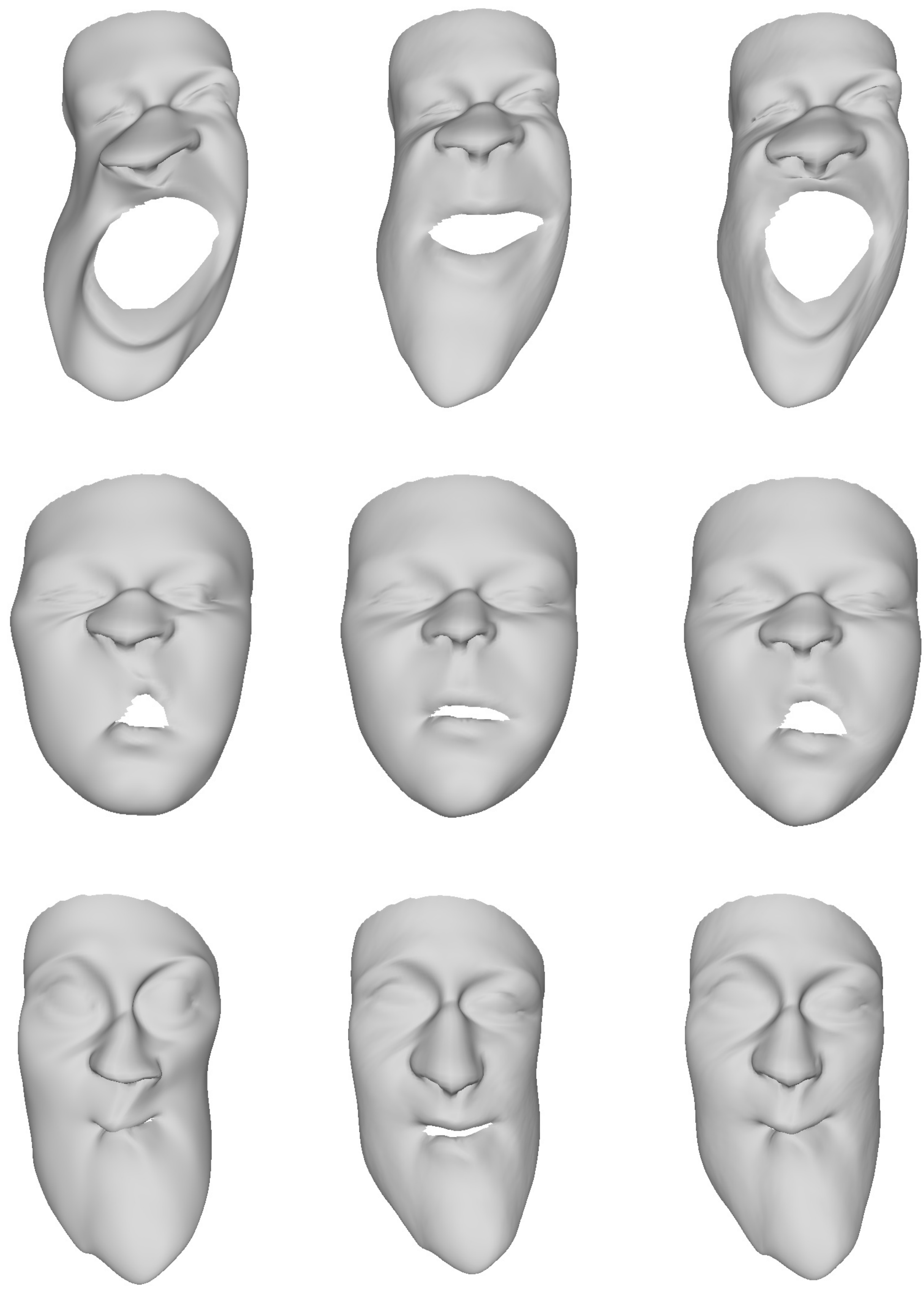}
\end{center}
\caption{Comparison result of VAE with and without expression constrained term $L_{\textrm{exp}}$. From left to right: input shape, reconstructed shape without $L_{\textrm{exp}}$, and reconstructed shape with $L_{\textrm{exp}}$.}
\label{fig:vae_noland}
\end{figure}

\begin{figure}[t]
\begin{center}
\includegraphics[width=1.0\linewidth]{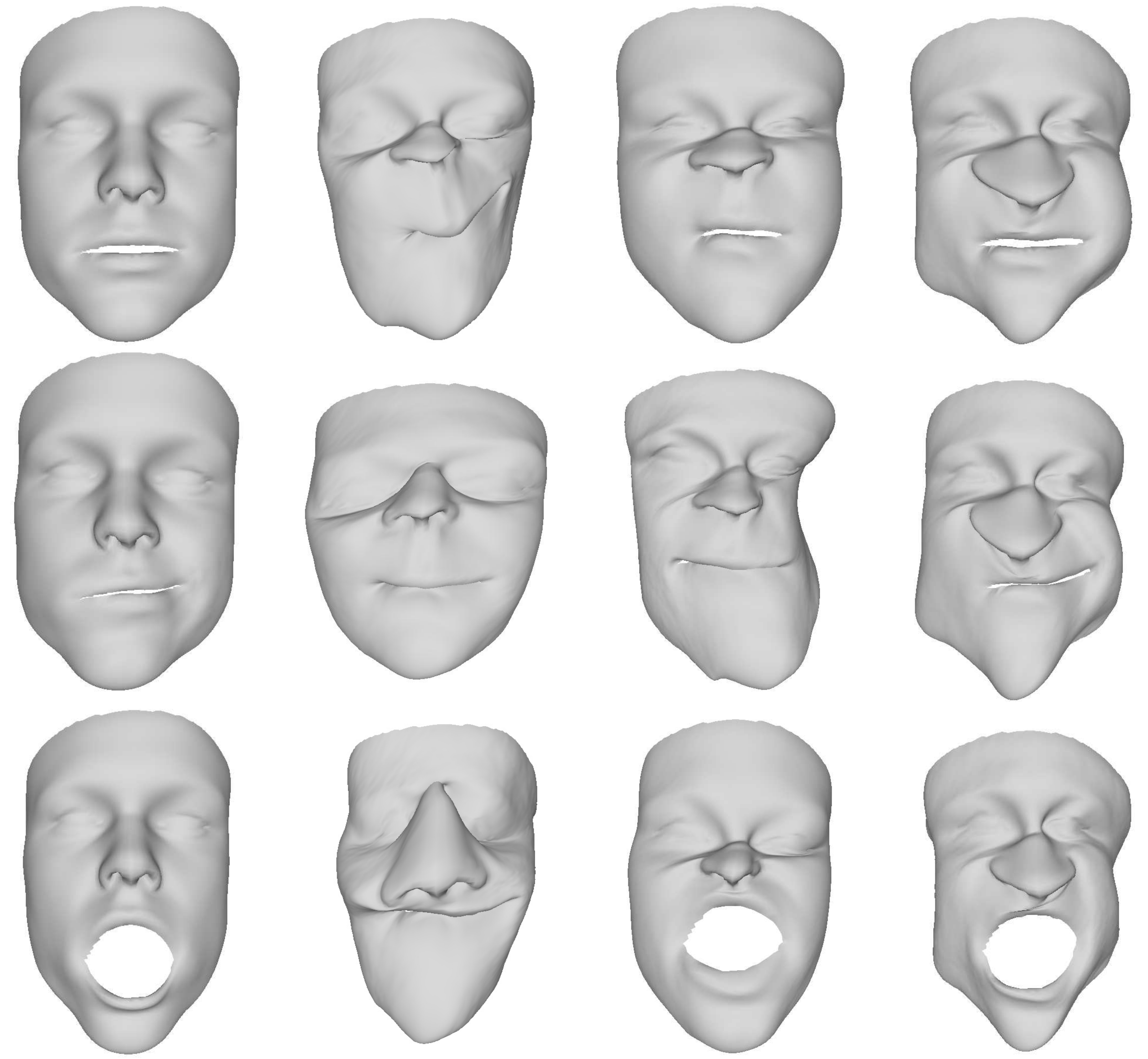}
\end{center}
\caption{Comparison result of CycleGAN with and without expression-preserving term $L_{\textrm{ang}}$ and identity-preserving term $L_{\textrm{pair}}$. From left to right: input shape, transferred shape without $L_{\textrm{ang}}$ and $L_{\textrm{pair}}$, transferred shape without $L_{\textrm{pair}}$, transferred shape with both terms.}
\label{fig:cycle_angid}
\end{figure}

\subsection{Ablation Study}
In this section, we analyze the necessity of some modules in our designed system. First, we compare our VAE-CycleGAN with CycleGAN that learns the translation between regular style and caricature style directly in the original 3D shape spaces, and the results are shown in Fig.~\ref{fig:novae}. It can be observed that different input face models are transformed into a similar messy 3D model without the embedded latent spaces. This is because our VAE-embedding helps constrain the translation within the reasonable face shape space. 

Second, we compare the designed VAE with the choice that excludes the landmark vertices based expression constrained term $L_{\textrm{exp}}$ in Fig.~\ref{fig:vae_noland}. It is shown that $L_{\textrm{exp}}$ could lead to more accurate expressions, especially on the mouth part. This is because that the landmark vertices are more important to convey the caricature styles, and we explicitly focus more on the alignment of landmark vertices by adding $L_{\textrm{exp}}$.

Finally, we show the necessity of the expression-preserving term $L_{\textrm{ang}}$ and identity-preserving term $L_{\textrm{pair}}$ by comparing the designed CycleGAN with choices that exclude these terms. The comparison results are shown in Fig.~\ref{fig:cycle_angid}. It can be observed that CycleGAN tends to learn a random mapping between latent spaces without $L_{\textrm{ang}}$ and $L_{\textrm{pair}}$ as shown in the second column, and it is hard to preserve identities without identity relevant loss terms $L_{\textrm{pair}}$ as shown in the third column.

\subsection{Comparison with Baseline Implementations}

Our proposed method is the first work to automatic 3D caricature modeling from 2D photos, and thus there is no existing algorithm for comparison. However, there exists some baseline implementations which could translate the input 2D photos to 3D caricatures. In the following, the comparisons between our method and these baseline implementations are given.

\noindent{\textbf{2D-Baseline.}} A baseline implementation is first to do caricature translation in the 2D domain and then reconstruct the corresponding 3D caricature models based on the 2D caricatures. Specifically, we get its corresponding 2D caricature by CariGANs~\cite{CaoLY18} with the input 2D photo, and then reconstruct the 3D caricature model by the method~\cite{WuZLZC18}.

\noindent{\textbf{3D-Baseline.}}  Another straight-forward method to do 3D caricature translation is to translate the geometry only with 3D landmarks, similar with 2D landmarks for 2D caricature generation in~\cite{CaoLY18}. Specifically, we train a network to translate the 3D landmarks of a regular 3D face to caricature style similar to the CariGeoGAN proposed in~\cite{CaoLY18}, and then deform the face shape based on exaggerated 3D landmarks using a Laplacian deformation algorithm~\cite{botsch2008linear}. 

\begin{figure*}[t]
\begin{center}
\includegraphics[width=1.0\linewidth]{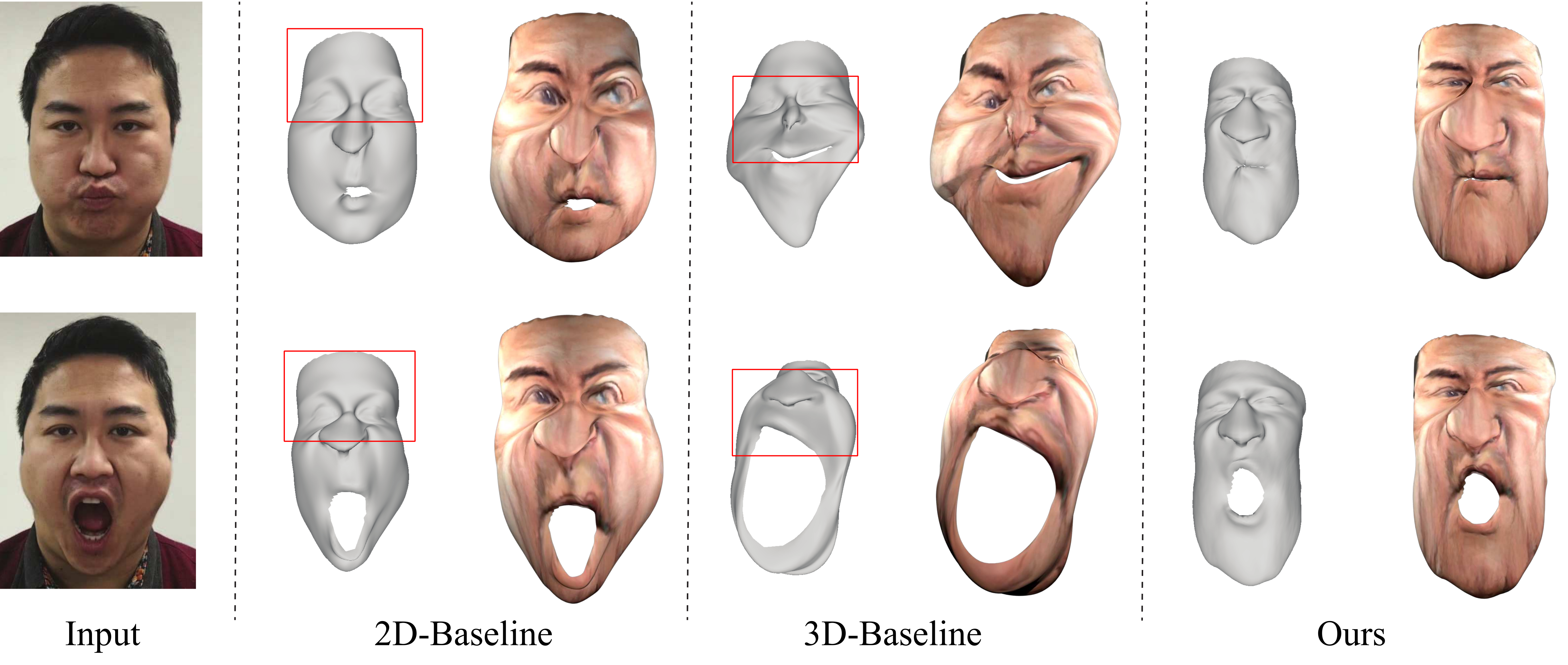}
\end{center}
\caption{Comparison result of our VAE-CycleGAN with 2D-Baseline and 3D-Baseline. From left to right: input frame, result of 2D-Baseline implementation, result of 3D-Baseline implementation, and our results. Red rectangles indicate 
inconsistent regions.}
\label{fig:deform_3dland}
\end{figure*}

\begin{figure}[t]
\begin{center}
\includegraphics[width=1.0\linewidth]{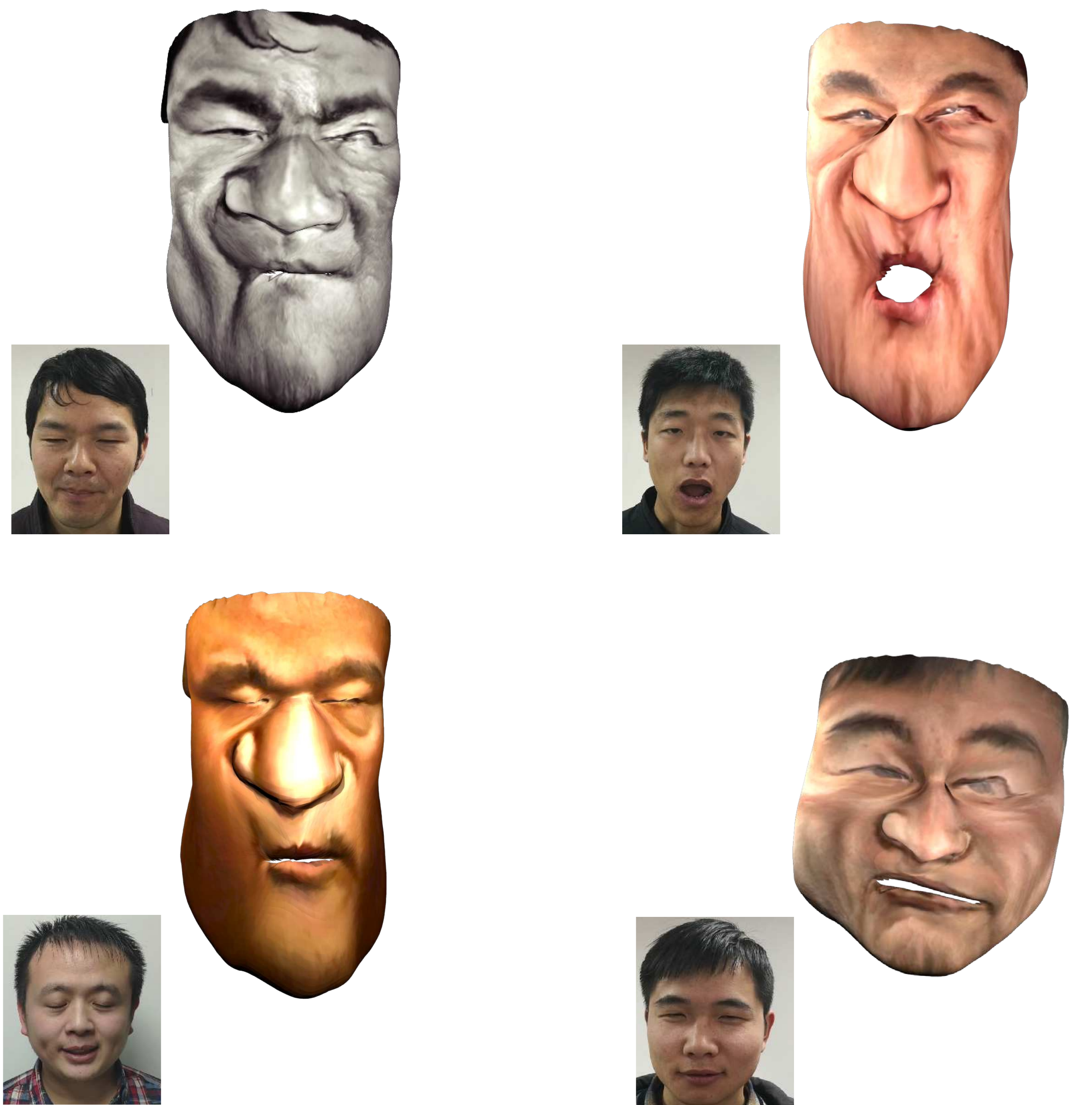}
\end{center}
\caption{More results by our method. The figures in the lower left corner are sampel frames of input video, and the central figures show the results by our method.}
\label{fig:moreresults}
\end{figure}

It takes about 40s for the 2D-baseline method to reconstruct a 3D caricature model from 2D caricature. For 3D-baseline method, it takes about extra 10ms to do laplacian deformation by using pre-computed sparse Cholesky factorization. We compare our VAE-CycleGAN with the two methods described above in Fig.~\ref{fig:deform_3dland}. It can be observed that our VAE-CycleGAN preserves the identity and expression better. 3D style transfer with only sparse landmarks might lead to some inconsistence for different expressions of the same person. More results by our method are shown in Fig.~\ref{fig:moreresults} and the supplementary video.

\subsection{Perceptual Study}
Since caricature is an artistic creation, its quality should be judged by the users. Therefore, we conducte a user study via an anonymous online web survey with 30 participants. Participants participating in the user study do not have experience in 3D modeling. We randomly pick 10 examples with the input 2D face videos and the results generated by three methods (ours, 2D-Baseline and 3D-Baseline) in random order. For the 2D-Baseline method, we registrate the reconstructed 3D caricature model with the reconstructed regular face at each frame to make the result be smooth with respect to pose.
 
\begin{figure}[t]
\begin{center}
\includegraphics[width=1.0\linewidth]{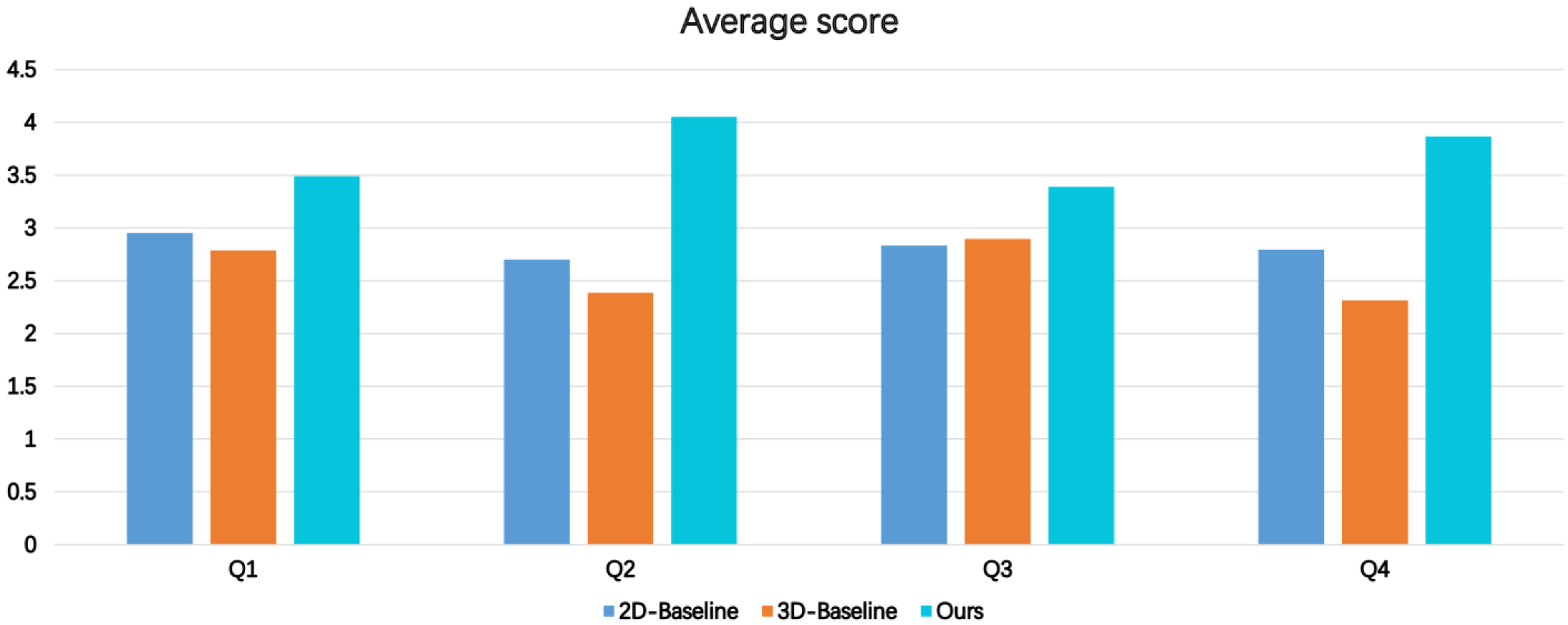}
\end{center}
\caption{Statistics of the user study results. Higher scores represent better results.}
\label{fig:user_study}
\end{figure}

For each example, every participant is requested to give scores to the following four questions: 
\begin{enumerate}[Q1.]
\item How is the expression consistency between the generated caricature sequence and the input face video?
\item How is the stability of the generated caricature sequence?
\item Does the exaggerated style of the generated caricature sequences match your expect?
\item Whether the identity of reconstructed caricature is preserved when expression changes? 
\end{enumerate}
We use numbers from 1 to 5 to represent the worst to the best. For each question, the higher the score, the better the result. The statistics of the user study are shown in Fig.~\ref{fig:user_study}, and the statistics indicate that our method outperforms other two baseline implementations on all the four aspects. The accompanying video shows the whole results of complete video sequences for one example.

\vspace{-5mm}

\section{Conclusion \& Discussion}
\label{sec:conclusion}

In this paper, we have presented the first approach for automatic 3D caricature modeling from 2D facial images. The static modeling part of the proposed method reconstructs a high-quality 3D face blendshape and a caricature style texture map for the user from multi-view facial images. In the face tracking stage, a well-designed 3D regular-to-caricature translation network has been proposed, which converts a dynamic 3D regular face sequence into a 3D caricature style and temporally-coherent sequence. More importantly, our proposed translation network ensures that the translated model maintains the same identity throughout the sequence, and the expression of each frame is consistent with the original model's expression. We have shown extensive experiments and user studies to demonstrate that our method outperforms baseline implementations in the aspect of visual quality, identity and expression preserving and computation speed.

\noindent\textbf{Limitations and future work.} Our approach still has some limitations. First, in order to meet the requirements of real-time computing, the 3D face blendshape is not further optimized during tracking process. This leads that our method can not reconstruct quite well for some expressions. One possible solution is to adopt the incremental learning framework proposed in~\cite{WuSS18}. Second, we tried to disentangle content and style on 3D face shapes by applying the MNUIT translation network~\cite{HuangLBK18}, but it failed. This results that our method can only output a 3D caricature model for each input. Third, current system only considers the facial part, not including the ears, and the part after the forehead. However, our system could exaggerate larger areas or even the entire head as long as the training set contains such data.

\section*{Acknowledgments}
The authors are supported by the National Natural Science Foundation of China (No. 61672481), and the Youth Innovation Promotion Association CAS (No. 2018495).

% or


%\appendices

%\section{}

%\ifCLASSOPTIONcompsoc
%\else
%\fi


%\newpage
\bibliographystyle{IEEEtran}
\bibliography{3DAliveCaricature}

\end{document}